\documentclass[letterpaper, 10 pt, conference]{ieeeconf}  % Comment this line out if you need a4paper

\IEEEoverridecommandlockouts                              % This command is only needed if 
\usepackage{amsthm}
\usepackage{graphics} % for pdf, bitmapped graphics files
\usepackage{epsfig} % for postscript graphics files
\usepackage{times} % assumes new font selection scheme installed
\usepackage{amsmath} % assumes amsmath package installed
\usepackage{amssymb}  % assumes amsmath package installed

\usepackage[maxbibnames=99,giveninits=true]{biblatex}
\usepackage{bm}
\usepackage{xcolor}
\usepackage{graphicx}
\usepackage{algorithm}
\usepackage[noend]{algpseudocode}
\usepackage{booktabs}
\usepackage{siunitx}% An awesome package for typesetting and manipulation numbers and units.
\usepackage{hyperref}% Used for insert the link

\usepackage{caption}% Better control over caption
\usepackage{lipsum}% Example text
\usepackage{multirow}
\usepackage{diagbox} 
\usepackage{subcaption}

\newcommand{\R}{\mathbb{R}}%commands for easy math notations
\newcommand{\io}{\iota_{\varepsilon}}
\newcommand{\casr}{\mathcal{R}^{\mathcal{I}_m,j}_{\varepsilon,C}}

\newcommand{\sigj}{\sigma_j}

\newtheorem{assumption}{\bf Assumption}
\newtheorem{theorem}{\bf Theorem}
\newtheorem{lemma}{\bf Lemma}

\title{\LARGE \bf
Emergence of Cascading Risk and Role of  Spatial Locations of Collisions  in Time-Delayed Platoon of  Vehicles  
}

\author{Guangyi Liu, Christoforos Somarakis, and Nader Motee% <-this % stops a space

\thanks{ G. Liu and N. Motee are with the Department of Mechanical Engineering and Mechanics, Lehigh University, Bethlehem, PA, 18015, USA. {\tt\small \{gliu,motee\}@lehigh.edu}.\endgraf
 C. Somarakis is with the Intelligent Systems Lab, Palo Alto Research Center - a Xerox Company, Palo Alto, CA, 94304. {\tt\small somarakis@parc.com.}
}
}

\begin{document}

\maketitle

% Replace this two for submission
\thispagestyle{plain}
\pagestyle{plain}

\begin{abstract}
We develop a framework to assess the risk of cascading collisions in a platoon of vehicles in the presence of exogenous noise and communication time-delay. The notion of Value-at-Risk (VaR) is adopted to quantify the risk of collision between vehicles in a pair conditioned on the knowledge of multiple previously occurred failures in the platoon. We show that the risk of cascading failures depends on the Laplacian spectrum of the underlying communication graph, time-delay, and noise statistics. Furthermore, we exploit the structure of several standard graphs to show how the risk profile depends on the magnitude and spatial location of the prior collisions (failures). Our theoretical findings are significant as they can be applied to designing safe platoons that minimize the risk of cascading failures. Our theoretical findings are supported by several simulations. 

%\guangyi{I suggest to connect this paper to the multi-vehicle accidents}

%\guangyi{If the page number exceeds 8, I'll fix it before the submission.}

\end{abstract}

%===============================================================================

\section{Introduction}
Uncertainties, originating from the fundamental laws of physics, prevail in every application in control systems. Even when a robust control system is designed to minimize its fragility, there is always a chance that the system will be driven into an undesired and even dangerous state in response to large exogenous shocks. The inherent stochasticity prevents one from precisely predicting the future state of the system. However, one can still assess the chance or the estimated cost of the system experiencing failure through notions of risk. The risk quantification plays a crucial role in almost all control systems as the evaluation of how ``risky'' a closed-loop system is will help both researchers and users to gain adequate knowledge about the safety and reliability of the system operation. 

The risk analysis in real-world applications can be traced from when \cite{korb2005risk} built the algorithmic risk measure for surgical robots to when \cite{terra2020safety} showed how the involvement of risk-mitigation in robot operations could increase safety. A systematic and analytical approach, which adopts the widely used financial tool value-at-risk (VaR) measure and the conditional-value-at-risk (CVaR) measure \cite{rockafellar2002conditional}, has recently been utilized to analyze the risk of certain events in the networked control system. Our previous works \cite{Somarakis2016g,Somarakis2017a, Somarakis2019g,Somarakis2020b, somarakis2018risk} have shown that the VaR measure is an effective tool to evaluate and analyze the safety in networked control systems, e.g., a platoon of vehicles, synchronous power networks, and rendezvous problems for a team of robots.

In this work, we consider the problem of autonomous vehicle platooning as a motivational application. This type of application is omnipresent in multi-agent systems \cite{grunewald2010energy,antonelli2004fault} or, more specifically, the car platooning \cite{tan1998demonstration}. An effective communication network is essential for closing a control loop, while the sensing and communication time-delays are inherent in all real-world applications. This phenomenon has also been observed and considered in some recent works about the vehicle platooning \cite{ali2015enhanced, verginis2015decentralized}. Hence, in this work, we consider platoon models that incorporate the effect of time-delay and noise in the dynamics of vehicles. 

% \guangyi{connect to multi-vehicle accidents here}

One fundamental design principle in a multi-vehicle platoon is to develop communication topologies that improve the platoon's reliability and safety. In \cite{Somarakis2020b}, the authors quantify the failure risk of one undesired event, e.g., inter-vehicle collision, and formulate several fundamental limits and trade-offs on the best achievable levels of risk. In a more recent work \cite{9683468}, we evaluate the collision risk of a  pair of vehicles conditioned on the event of exactly one prior collision in the platoon, where it is shown that the risk of cascading failures is not only dependent on the network parameters (e.g., Laplacian eigen-spectrum, time delay) and noise statistics, but also affected by the relation between the existing collision and the pair of the interest. In this work, which is along with our previous works \cite{Somarakis2016g,Somarakis2017a, Somarakis2019g,Somarakis2020b, somarakis2018risk} to analyze the risk of collisions in a platoon of vehicles, we propose a method that allows us to calculate cascading risk of collision between vehicles in a pair conditioned on knowledge of an arbitrary number of prior collisions. Our goal is to explore how design paradigms will change while considering all possible collision scenarios.

Our distinct contributions are twofold. First, we 
derive explicit formulas for the cascading risk of collision subject to a series of known malfunctioned vehicles (failures). Second, we focus on exploring how the topology of a communication graph, the scale, and the spatial distribution of the prior failures will affect the risk of cascading collision. We obtain closed-form risk representation for some particular graph topologies by considering various scenarios for the spatial distribution of prior collisions. We should highlight that the behavior of cascading risk is significantly more complicated than the risk of a single event, and that may prevent one from drawing general conclusions about risk.  

% {\BC Make sure to clearly state the differences between this work and our earlier work on this topic}
% \guangyi{Will do by tonight.}

%The rest of the paper is organized as follows. In \S \ref{sec:prelims} we present a few key results on the steady-state behavior and statistics of the closed-loop autonomous vehicle platoon. Then, we shape the risk formula of generalized cascading failures in \S \ref{sec:group_risk}, which constitutes the major contribution of this work. Next, the closed-form risk formula of cascading failures in a complete communication graph is examined in \S \ref{specialgraph}. Finally, in \S \ref{sec:case-study}, detailed simulations are present to reveal the complexity of the risk of multi-vehicle accidents.

% \guangyi{Connect the multi-vehicle accident with the cascading failure.}

%===============================================================================
\section{Preliminaries}
The $n-$dimensional Euclidean space  is denoted by $\R^n$ and $\mathbb{R}^n_{+}$ represents the positive orthant of $\R^n$. We denote the vector of all ones by $\bm{1}_n = [1, \dots, 1]^T$. The set of standard Euclidean basis for  $\mathbb{R}^{n}$ is represented by $\{\bm{e}_1, \dots, \bm{e}_n\}$ and  $\tilde{\bm{e}}_i := \bm{e}_{i+1} - \bm{e}_{i}$ for all $i = 1, \dots, n-1$. 

{\it Algebraic Graph Theory:} A weighted graph is defined by $\mathcal{G} = (\mathcal{V}, \mathcal{E}, \omega)$, where $\mathcal{V}$ is the set of nodes, $\mathcal{E}$ is the set of edges (feedback links), and $\omega: \mathcal{V} \times \mathcal{V} \rightarrow \mathbb{R}_{+}$ is the weight function that assigns a non-negative number (feedback gain) to every link. Two nodes are directly connected if and only if $(i,j) \in \mathcal{E}$.

\begin{assumption}  \label{asp:connected}
    Every graph in this paper is connected. In addition, for every $i,j \in \mathcal{V}$, the following properties hold:
    \begin{itemize}
        \item $\omega(i,j) > 0$ if and only if $(i,j) \in \mathcal{E}$.
        \item $\omega(i,j) = \omega(j,i)$, i.e., links are undirected.
        \item $\omega(i,i) = 0$, i.e., links are simple.
    \end{itemize}
    
\end{assumption}

The Laplacian matrix of $\mathcal{G}$ is a $n \times n$ matrix $L = [l_{ij}]$ with elements
\[
    l_{ij}:=\begin{cases}
        \; -k_{i,j}  &\text{if } \; i \neq j \\
        \; k_{i,1} + \ldots + k_{i,n}  &\text{if } \; i = j 
    \end{cases},
\]
where $k_{i,j} := \omega(i,j)$. Laplacian matrix of a graph is symmetric and positive semi-definite \cite{van2010graph}. Assumption \ref{asp:connected} implies the smallest Laplacian eigenvalue is zero with algebraic multiplicity one. The spectrum of $L$ can be ordered as 
$$
    0 = \lambda_1 < \lambda_2 \leq \dots \leq \lambda_n.
$$
The eigenvector of $L$ corresponding to $\lambda_k$ is denoted by $\bm{q}_{k}$. By letting $Q = [\bm{q}_{1} | \dots | \bm{q}_{n}]$, it follows that $L = Q \Lambda Q^T$ with $\Lambda = \text{diag}[0, \lambda_2, \dots, \lambda_n]$. We normalize the Laplacian eigenvectors such that $Q$ becomes an orthogonal matrix, i.e., $Q^T Q = Q Q^T = I_{n}$ with $\bm{q}_1 = \frac{1}{\sqrt{n}} \bm{1}_n$. 

{\it Probability Theory:} Let $\mathcal{L}^{2}(\mathbb{R}^{q})$ be the set of all $\R^q-$valued random vectors $\bm{z} = [z^{(1)}, \dots ,z^{(q)}]^T$ of a probability space $(\Omega, \mathcal{F}, \mathbb{P})$ with finite second moments. A normal random variable $\bm{y} \in \mathbb{R}^{q}$ with mean $\bm{\mu} \in \mathbb{R}^{q}$ and $q \times q$ covariance matrix $\Sigma$ is represented by $\bm{y} \sim \mathcal{N}(\bm{\mu}, \Sigma)$. The error function $\text{erf}:\R \rightarrow (-1,1)$ is
$
\text{erf} (x) = \frac{2}{\sqrt{\pi}} \int_{0}^{x} e^{-t^2} \text{d} t,
$
which is invertible on its range as $\text{erf}^{-1} (x)$. We employ standard notation $\text{d} \bm{\xi}_t$ for the formulation of stochastic differential equations.

%%%%%%%%%%%%%%%%%%%%%%%%%%%%%%%%%%%%%%%%%%%%%%%%%%%%%%%%%%%%%%%%%%%%%%%%%%%%%%%%%
\section{Problem Statement}\label{problemstatement}
\begin{figure}[t]
    \centering
	\includegraphics[width=\linewidth]{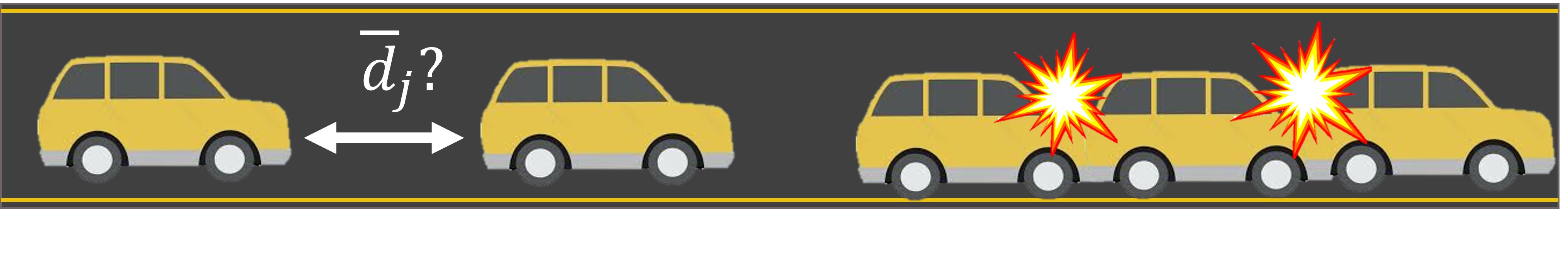}
	\caption{Schematics of platoon ensemble of autonomous vehicles. We assume some vehicles have collided and we aim to find the risk of cascading collision in the $j$'th pair.} 
	\label{fig:platoon}
\end{figure}

Suppose that a finite number of vehicles $\mathcal{V} = \{1, \ldots, n\}$ form a platoon along the horizontal axis. Vehicles are labeled in descending order, where the $n$’th vehicle is assumed to be the leading vehicle in the platoon. The $i$’th vehicle’s state is determined by $[x_t^{(i)}, v_t^{(i)}]^T$ , where $x_t^{(i)}$ is the position and $v_t^{(i)}$ is the velocity of vehicle $i\in \mathcal{V}$ at time $t$. The dynamics of the $i$’th vehicle evolves according to stochastic differential equation
\begin{equation} \label{eq:dyn_vehicle}
    \begin{aligned}
        d x^{(i)}_{t} &= v^{(i)}_{t} dt\\
        d v^{(i)}_{t} &= u^{(i)}_{t} dt + g\, d \xi^{(i)}_{t}
    \end{aligned}
\end{equation}
where $u^{(i)}_t \in \R$ is the control input at time $t$. The term $g\, d \xi^{(i)}_{t}$  represents the white noise generator that  models the uncertainty diffused in the system. It is assumed that noise acts on every vehicle additively and independently from the other vehicles’ noises. The noise magnitude is represented by the diffusion $g \neq 0$, which is assumed to be identical for all $i \in \mathcal{V}$. The control objectives for the platoon are to guarantee the following two global behaviors: (i)  the relative distance between every two consecutive vehicles converges to a prescribed distance; and (ii) the platoon of vehicles attain the same constant velocity in steady state. To incorporate deficiencies of communication network, we assume all vehicles experience an identical and constant
% \footnote{This assumption has been widely used by other researchers as it allows analytical derivations of formulas. It is a common practice in vehicles labs to use identical communication modules for all vehicles, which results in a uniform communication time-delay. Moreover, in other related applications such as heading alignments, rendezvous in space, and velocity control of vehicles in indoor labs, where a motion capture (MoCap) system is utilized to localize the vehicles \cite{lupashin2014platform,michael2010grasp}, all vehicles experience an identical time-delay to access data through the MoCap system.} 
time-delay $\tau \in \mathbb{R}_{+}$. It is known from \cite{yu2010some} that the following feedback control law can achieve the platooning objectives 
\begin{equation} \label{eq:feedback}
    \begin{aligned}
        u^{(i)}_{t} = \sum_{j=1}^{n} & ~k_{i,j} \Big( v^{(j)}_{t - \tau} - v^{(i)}_{t - \tau}\Big)\\
        & + ~\beta \sum_{j=1}^{n} k_{i,j} \Big( x^{(j)}_{t - \tau} - x^{(i)}_{t - \tau} - (j-i) d \Big).
    \end{aligned}
\end{equation}
The parameter $\beta \in \R_{++}$ regulates the effect of the relative positions and the velocities. By defining the vector of positions, velocities, and noise inputs as $\bm{x}_t = [x_{t}^{(1)}, \dots, x_{t}^{(n)}]^T$, $\bm{v}_t = [v_{t}^{(1)}, \dots, v_{t}^{(n)}]^T$ and $\bm{\xi}_t = [\xi_{t}^{(1)}, \dots, \xi_{t}^{(n)}]^T$, respectively, we denote the target distance vector as $\bm{r} = [d,2d,\dots,nd]^T$. Using the above control input, we represent the closed-loop dynamics as an initial value problem
\begin{equation}\label{eq:dyn}
    \begin{aligned}
        d\bm{x}_t &= \bm{v}_t dt,\\
        d\bm{v}_t &= -L \bm{v}_{t-\tau} dt - \beta L (\bm{x}_{t-\tau} - \bm{r}) dt+ g d \bm{\xi}_t,
    \end{aligned}
\end{equation}
for all $t\geq 0$ and given deterministic initial function of $\bm \phi^{\bm{x}}_t$ and $\bm \phi^{\bm{v}}_t \in \R^n$ for $t \in [-\tau, 0]$. It is well known that \eqref{eq:dyn} generates a well-posed stochastic process $\{(\bm{x}_t, \bm{v}_t)\}_{t\geq -\tau}$, \cite{mohammed1984stochastic}.

The {\it problem} is to quantify the collision risk of one particular pair of vehicles conditioned on  the knowledge of knowing a subset of pairs  have already collided. We refer to this quantity as cascading collisions (failures) risk and our goal is to characterize its value as a function of the underlying communication graph topology, time-delay, and statistics of noise. To this end, we will develop a general framework to assess cascading risk of collisions using the steady-state statistics of the closed-loop system \eqref{eq:dyn_vehicle} and \eqref{eq:dyn}. 

%%%%% 
%We should cite \cite{9683468} in  Introduction section. 
%%%%%%%%%%%%%%

% The rest of the paper is organized as follows. In \S \ref{sec:prelims} we present a few key results on the steady-state behavior and statistics of the closed-loop mobile robot platooning  system. Using the above results, we shape the risk formula of group cascading failures in the \S \ref{sec:group_risk}, which constitutes the major contribution of this work. \S \ref{limits} is devoted to design challenges for risk minimization with general remarks and the case study of a complete communication graph. Special graph topologies are examined in \S \ref{specialgraph} where we investigate theoretically and by simulation the complexity of group cascading failure phenomena. All proofs of our theoretical results are presented in the appendix. {\BC rewrite}

%%%%%%%%%%%%%%%%%%%%%%%%%%%%%%%%%%%%%%%%%%%%%%%%%%%%%%%%%%%%%%%%%%%%%%%%%%%%%%%%%%%%%
\section{Preliminary Results}\label{sec:prelims}
To evaluate the risk of cascading collisions in a platoon, we briefly review some necessary concepts and results \cite{Somarakis2020b}.   

\subsection{Platooning State and Stability Conditions}
Keeping a safe constant distance from each other while traversing with a constant velocity is commonly referred to as the target (or consensus) state in a platoon \cite{klanvcar2011control,bamieh2012coherence}. We say that the group of vehicles with dynamics \eqref{eq:dyn} forms a platoon if 
\begin{equation*}
    \lim_{t \rightarrow \infty} |v_{t}^{(i)} - v_{t}^{(j)}| = 0 \,\text{and}\, \lim_{t \rightarrow \infty} |x_{t}^{(i)} - x_{t}^{(j)} - (i-j) d| = 0
\end{equation*}
for all $i,j \in \mathcal{V}$ and all initial functions. It is known \cite{yu2010some,Somarakis2020b} that the deterministic time-delayed network of vehicles  will converge and form a platoon if and only if 
$
    (\lambda_i \tau, \beta \tau ) \in S
$
for all $i=2,\ldots,n$ where
\begin{align*}
    S = \bigg\{(s_1,s_2) \in \R^2 \bigg| s_1 \in \left(0, \frac{\pi}{2}\right), s_2 \in \left(0, \frac{a}{\tan(a)}\right) \\\text{for } a \in \left(0,\frac{\pi}{2}\right) \text{ the solution of } a\sin(a) = s_1 \bigg\}.
\end{align*} 
Throughout the paper, we assume that the underlying deterministic network of vehicles forms a platoon.

\subsection{Steady-State Inter-vehicle Distance}   \label{statistics}
In present of stochastic exogenous noise, inter-vehicle failures are mainly due to large deviations in inter-vehicle distances from a desired safe distance \cite{ali2015enhanced}. Let us denote   
the steady-state value of the relative distance between vehicles $i$ and $i+1$ by $\bar{d}_i = \bar{x}_{t}^{(i+1)} - \bar{x}_{t}^{(i)}$ and $\bm{\bar{d}}=[\bar{d}_1, \ldots, \bar{d}_{n-1}]^T$. 
% \chris{this is a different notation from above (Eq. (3) ). Am i missing sth ? } \guangyi{$\bm{r}$ now denotes the target distances. Now it should look better} \chris{ makes sense. }
\begin{lemma}\label{lem:d_steady}
    \cite{9683468} Suppose that the network of vehicles forms a platoon. Then, random variable $\bar{\bm{d}}$ has a multi-variate normal distribution in $\R^{n-1}$ that is given by
    \begin{align*}  
        \bm{\bar{d}} \sim \mathcal{N}(d \bm{1}_n, \Sigma),
    \end{align*}
    where the elements of the covariance matrix $\Sigma = [\sigma_{ij}]$ are
    \begin{equation} \label{eq:sigma_d}
    \begin{aligned}
        \sigma_{ij} = 
        g^2 \frac{\tau^3}{2\pi} \sum_{k=2}^{n} \big(\tilde{\bm{e}}_{i}^T \bm{q}_k \big) \big(\tilde{\bm{e}}_{j}^T \bm{q}_k \big) f(\lambda_k \tau, \beta \tau),
    \end{aligned}
    \end{equation}
    for all $i,j=1,\ldots,n$ and 
    $$
    f(s_1, s_2) = \int_{\R} \frac{\text{d}\,r}{(s_1s_2 - r^2 \cos(r))^2 + r^2 (s_1-r \sin(r))^2}.
    $$
\end{lemma}
For the simplicity of notations, we use  $\sigma_i^2$ instead of $\sigma_{ii}$.
\subsection{Risk of a Single Collision}
To quantify the uncertainty level encapsulated in the relative distances between the vehicles, we employ the notion of Value-at-Risk  \cite{Follmer2016,rockafellar2000optimization,Somarakis2016g,Somarakis2020b}. The VaR indicates the chance of a random variable landing inside an undesirable set of values, i.e., a near-collision situation. The set of undesirable values is referred to a systemic set, which are denoted as $C \subset \R$. In probability space $(\Omega, \mathcal{F}, \mathbb{P})$, the set of systemic events of random variable $y: \Omega \rightarrow \R$ is define as $\{ \omega \in \Omega ~|~y(\omega) \in C\}$. We define a collections of super-sets $\{C_{\delta}~|~\delta \in [0,\infty]\}$ of $C$ that satisfy the following conditions for any sequence $\{\delta_n\}^{\infty}_{n=1}$ with property $\lim_{n \rightarrow \infty} \delta_n = \infty$
\begin{itemize}
    \item $C_{\delta_1} \subset C_{\delta_2}$ when $\delta_1 > \delta_2$.
    \item $\lim_{n \rightarrow \infty} C_{\delta_n} = \bigcap_{n=1}^{\infty} C_{\delta_n} = C$.
\end{itemize}

In practice, one can tailor the super-sets to cover a suitable neighborhood of $C$ to characterize alarm zones as a random variable approaches $C$. For a given $\delta >0$, the chance of $\{ y \in C_{\delta}\}$ indicates how close $y$ can get to $C$ in probability. For a given design parameter $\varepsilon \in (0,1)$, the VaR measure $\mathcal{R}_{\varepsilon} : \mathcal{F} \rightarrow \R_{+}$ is defined by
\begin{equation*}
    \mathcal{R}_{\varepsilon} := \inf \left\{ \delta>0 ~\bigg|~ \mathbb{P} \left\{y \in C_{\delta} \right\} < \varepsilon \right\},
\end{equation*}
where a smaller $\varepsilon$ indicates a  higher level of confidence on random variable $y$ to stay away from $C_\delta$. Let us elaborate and interpret what typical values of $\mathcal{R}_{\varepsilon}$ imply. The case $\mathcal{R}_{\varepsilon}=0$ signifies that the probability of observing $y$ dangerously close to $C$ is less than $\varepsilon$. We have $\mathcal{R}_{\varepsilon} > 0$ iff $y \in C_{\delta}$ for some $\delta >0$ (in fact, $\delta >  \mathcal{R}_{\varepsilon}$) with probability greater than $\varepsilon$. The extreme case $\mathcal{R}_{\varepsilon}=\infty$ indicates that the event that $y$ is to be found in $C$ is assigned with a probability greater than $\varepsilon$. In addition to several interesting properties (see for instance \cite{artzner1997thinking,artzner1999coherent,Somarakis2020b}), the risk measure is non-increasing with $\varepsilon$. We refer to Fig. \ref{fig:risk-set} for an illustration.

%%%%%%%%%%%%%%%%%%%%%%%%%%%%%%%%%%%%%%%%%%%%%%%%%%%%%%%%%%%%%%%%%%%%%%%%%%%%%%%%%%%%%
\section{Risk of Cascading  Collisions}  \label{sec:group_risk}
\begin{figure}[t]
    \centering
	\includegraphics[width=\linewidth]{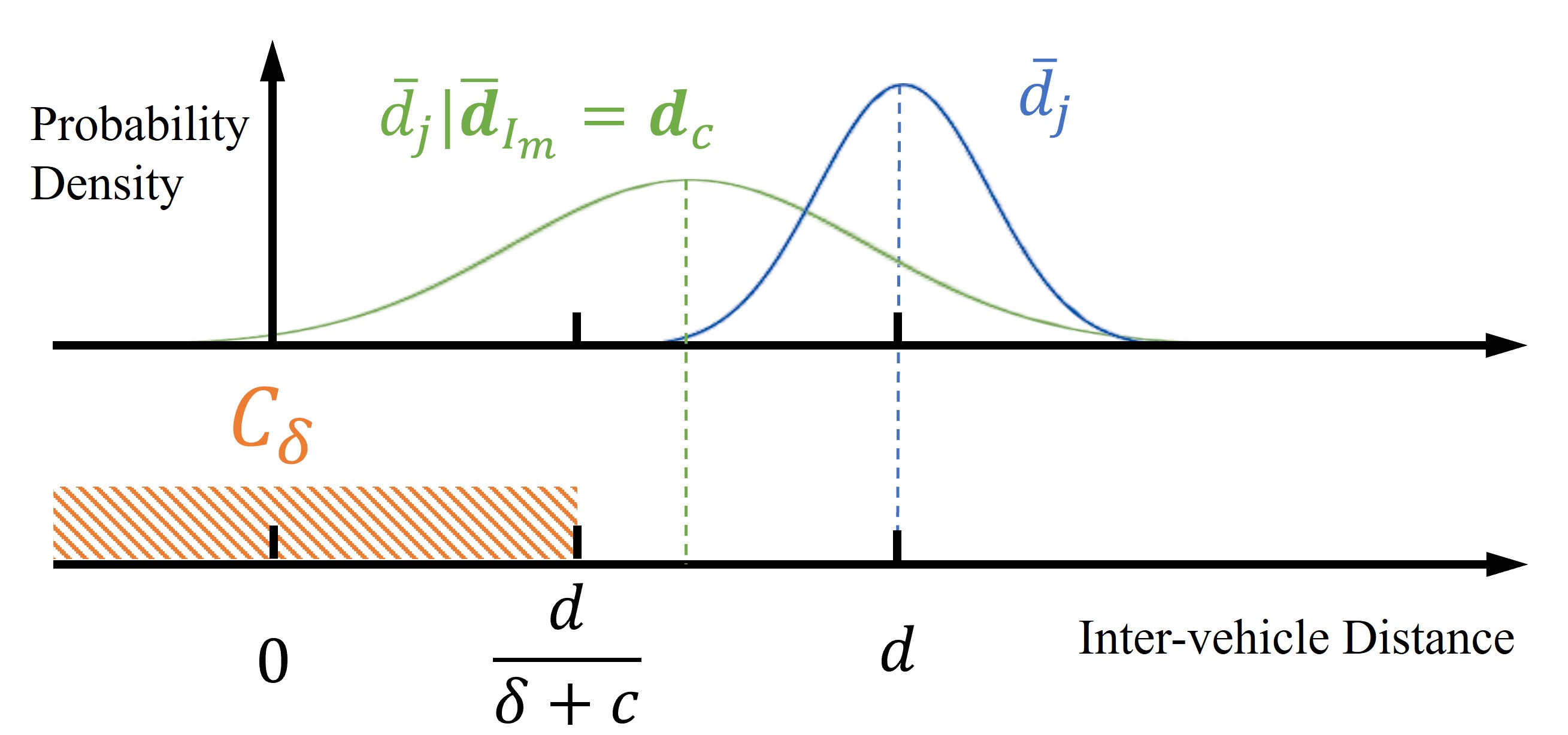}
	\caption{This figure depicts the concept of steady-state distance $\bar{d}_j$, conditional distance $\bar{d}_j | \bm{\bar{d}}_{\mathcal{I}_m} = \bm{d}_{c}$ and the collision set $C_{\delta}$. }
	\label{fig:risk-set}
\end{figure}

% \nader{Let's use 'collision' instead of 'failure' when we are  obtaining our results and theorems. Also, use 'platoon' instead of 'system' or 'networked control system'}

% \guangyi{I'll fixed most of them, and for the rest I'm not confident to change. Please feel free to change the rest of them.}

In \cite{9683468}, we consider the collision risk of a particular pair conditioned on the knowledge  of  {\it only one} prior collision. However, in platoons the chances of encountering {\it multiple} collisions are not negligible and it is desirable to design platoons that are robust w.r.t cascade of multiple collisions. Thus, as an organic outgrowth of \cite{9683468, Somarakis2020b}, in this section, we develop tools to assess cascading risk conditioned on an arbitrary number of prior collisions.

Let us consider the collision risk between the vehicles in the $j$'th pair and assume that pairs with ordered indices $\mathcal{I}_m =\{i_1, \cdots, i_m\}$ for some $m < n$ have already experienced collisions, where $j \notin \mathcal{I}_m$ and $i_1<i_2<...<i_m$. To evaluate the effect of the prior collisions on the $j$'th pair, we form the $(m+1) \times (m+1)$ block covariance matrix 
\begin{equation}    \label{eq:new-sig}
    \tilde{\Sigma} = \begin{pmatrix}\,
       \tilde{\Sigma}_{11} &\tilde{\Sigma}_{12}\\
       \tilde{\Sigma}_{21} &\tilde{\Sigma}_{22}\,
    \end{pmatrix}
\end{equation}
using \eqref{eq:sigma_d}, where
$
\tilde{\Sigma}_{11} = \sigma_{j}^2, \, \tilde{\Sigma}_{12} = \tilde{\Sigma}_{21}^T = [\sigma_{j i_1},...,\sigma_{j i_m}]$, and  $\tilde{\Sigma}_{22} = [\sigma_{k_1 k_2}]_{k_1, k_2 \in \mathcal{I}_m} \in \R^{m \times m}$. Suppose that the states of the collided pairs are represented by vector $\bm{d}_{c} = [d_{c_1}, ... , d_{c_m}]^T$. For instance, when the vehicles in the $i$'th pair are detached, one may set $d_{c_i}$ to $2d$, and when they are about to collide one may set $d_{c_i}$ to   $0.1 d$ or even $0$. We calculate the conditional probability distribution of $\bar{d}_j$ given 
$\bm{\bar{d}}_{\mathcal{I}_m}  = \bm{d}_{c}$ 
in a multi-variate normal distribution, where $\bm{\bar{d}}_{\mathcal{I}_m}=[\bar{d}_{i_1}, ... , \bar{d}_{i_m}]^T$.

\begin{lemma}   \label{lem:multi-condition}
    Suppose that the assumptions of Lemma \ref{lem:d_steady} hold. Then, the conditional distribution of $\bar{d}_j$ given $\bm{\bar{d}}_{\mathcal{I}_m} = \bm{d}_{c}$ is a normal distribution  $\mathcal{N}(\tilde{\mu}, \tilde{\sigma})$ with
    \begin{equation}
        \tilde{\mu} = d + \tilde{\Sigma}_{12} \, \tilde{\Sigma}_{22}^{-1}(\bm{d}_c -  d\bm{1}_m)
    \end{equation}
    and 
    \begin{equation}
        \tilde{\sigma}^2 = \tilde{\Sigma}_{11} - \tilde{\Sigma}_{12} \, \tilde{\Sigma}_{22}^{-1} \, \tilde{\Sigma}_{21},
    \end{equation}
    where the sub-blocks $\tilde{\Sigma}_{11}, ..., \tilde{\Sigma}_{22}$ are defined in \eqref{eq:new-sig}.
\end{lemma}

The proof of this lemma is eliminated due to page-limit; the main lines of proof relies on Lemma and calculation of conditional distribution of a multi-variate normal random variable \cite{tong2012multivariate}. 

We characterize the cascading risk of the $j$'th pair  by 
\begin{align*}
    \Big\{ \bar{d}_{j} \in C_{\delta} \;\big|\; \bm{\bar{d}}_{\mathcal{I}_m} = \bm{d}_{c}  \Big\} \text{ with } C_{\delta}:=\Big( -\infty, \frac{d}{\delta+c} \Big)
\end{align*}
for $\delta \in [0,\infty]$ and design parameter $c \geq 1$.  The systemic set of collision is $C=(-\infty,0)$. A visualization of this setup is depicted in Fig. \ref{fig:risk-set}. The value-at-risk measure for the cascading collision event is defined as
\begin{align}\label{eq:var}
    \mathcal{R}^{\mathcal{I}_m,j}_{\varepsilon,C} := \inf \bigg\{ \delta > 0 \;\bigg|\; \mathbb{P} \left\{ \bar{d}_{j} \in C_{\delta} \;\big|\; \bm{\bar{d}}_{\mathcal{I}_m} = \bm{d}_{c} \right\} < \varepsilon \bigg\}
\end{align}
with the confidence level $\varepsilon \in (0,1)$, where $j \notin \mathcal{I}_m$. The next theorem presents a closed-form representation of the  cascading collision risk \footnote{The risk of other types of failures, e.g., detachments \cite{Somarakis2020b}, can be derived using similar lines of analysis.}.

\begin{theorem}     \label{thm:1}
   Suppose that the vehicles form a platoon and it is known that the $i_1, \cdots, i_m$'th pairs of vehicles have already collided. The risk of cascading  collision of the $j$'th pair is 
    \begin{equation}    \label{eq:thm-1}
        \mathcal{R}^{\mathcal{I}_m,j}_{\varepsilon,C}=\begin{cases}
            0, &\text{if} ~ \frac{d-c\tilde{\mu}}{\sqrt{2}\tilde{\sigma} c} \leq \iota_{\varepsilon}\\
            \dfrac{d}{\sqrt{2} \io \tilde{\sigma} + \tilde{\mu}} - c, &\text{if} ~ \iota_{\varepsilon} \in \big(\frac{-\tilde{\mu}}{\sqrt{2} \tilde{\sigma}}, \frac{d-c\tilde{\mu}}{\sqrt{2}\tilde{\sigma} c} \big) \\
            \infty, &\text{if} ~ \frac{-\tilde{\mu}}{\sqrt{2} \tilde{\sigma}} \geq  \iota_{\varepsilon}
            \end{cases},
    \end{equation}
    where $\tilde{\mu}$ and $\tilde{\sigma}$ are given in Lemma \ref{lem:multi-condition}, $\io = \text{erf}^{-1}(2\varepsilon - 1)$.
\end{theorem}

\begin{proof}
    Following Lemma \ref{lem:multi-condition}, one can rewrite \eqref{eq:var} as 
    \begin{align*}
        \inf \left\{ \; \delta > 0 \; \Big| \; \int_{-\infty}^{ \frac{d - \tilde{\mu} (\delta+c)}{\sqrt{2} \tilde{\sigma} (\delta+c)} } \exp{\big( -t^2 \big)} \; dt < \sqrt{\pi} \varepsilon\;\right\}.
    \end{align*}
    We need to consider three cases. Let us first consider $\casr = 0$, which is equivalent to 
    \begin{align*}
        \int_{-\infty}^{ \frac{d - \tilde{\mu} c}{\sqrt{2} \tilde{\sigma} c} } \exp{\big( -t^2 \big)} \; dt < \sqrt{\pi} \varepsilon ~ \Leftrightarrow  ~ \text{erf}(\frac{d - c  \tilde{\mu}}{\sqrt{2} \tilde{\sigma} c})< 2 \varepsilon-1,
    \end{align*}
one can then use the inverse error function $\text{erf}^{-1}(\cdot)$ to obtain the branch condition. The similar argument works for the case $\delta \rightarrow \infty$, where one can use the term $\frac{- \tilde{\mu}}{\sqrt{2} \tilde{\sigma}}$ to compute the branch condition. For the intermediate branch, we obtain the positive solution, i.e., $\delta >0$,  by solving 
    $$
    \int_{-\infty}^{\frac{d - \tilde{\mu} (\delta+c)}{\sqrt{2} \tilde{\sigma} (\delta+c)}} e^{-t^2}\,dt=\sqrt{\pi}\varepsilon, 
    $$
    where the uniqueness of this solution is guaranteed by the monotonicity of $\frac{d - \tilde{\mu} (\delta+c)}{\sqrt{2} \tilde{\sigma} (\delta+c)}$.
\end{proof}

\begin{figure}[t]
    \centering
	\includegraphics[width=0.9\linewidth]{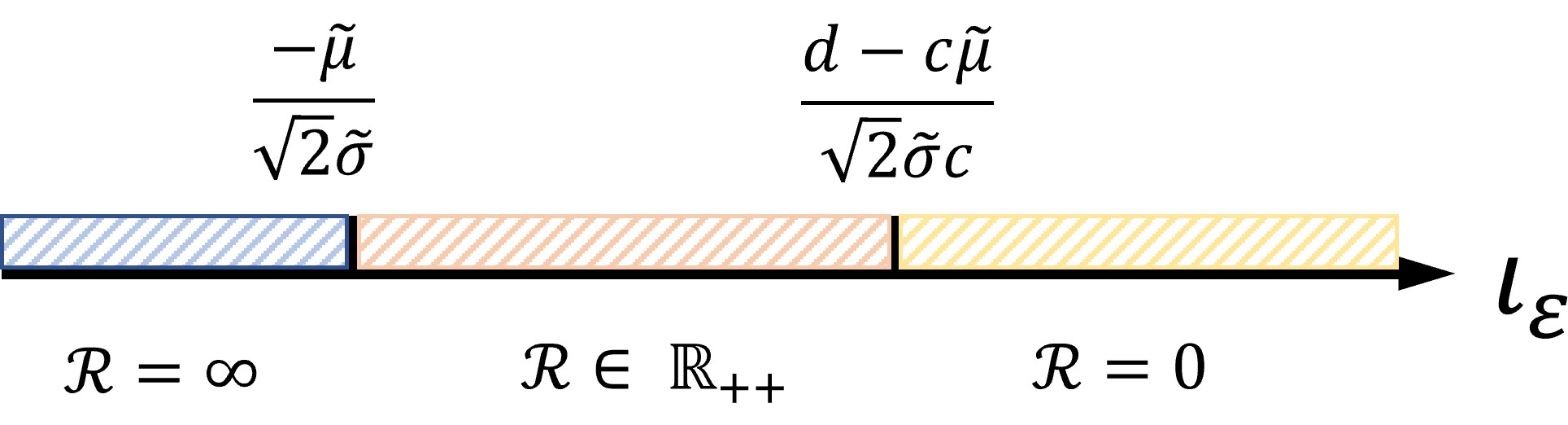}
	\caption{The risk partition on the axis of $\io = \text{erf}^{-1}(2\varepsilon - 1
    )$. }
	\label{fig:iota}
\end{figure}

Based on the chosen confidence level $(1 - \varepsilon)$, the cascading risk falls into three categories. In the first scenario, if the confidence level $(1 - \varepsilon)$ is too low, we have  $\mathbb{P}\{\bar{d}_j \in C_{\delta} \;\big|\; \bm{\bar{d}}_{\mathcal{I}_m} = \bm{d}_{c} \} < \mathbb{P}\{\bar{d}_j \in C_{0} \;\big|\; \bm{\bar{d}}_{\mathcal{I}_m} = \bm{d}_{c} \} \leq \varepsilon$ for every $\delta > 0$. Hence, there is {\it no risk} of having further collision with the confidence level $(1-\varepsilon)$; see the yellow area in Fig. \ref{fig:iota}. The second case indicates that if the confidence level $(1 - \varepsilon)$ is too high, the risk of having the cascading collision will be {\it infinitely large} and no $\delta$  can bound the value of $\mathbb{P}\{\bar{d}_j \in C_{\delta} \;\big|\; \bm{\bar{d}}_{\mathcal{I}_m} = \bm{d}_{c} \}$ with $\varepsilon$; see the blue area in Fig. \ref{fig:iota}. The third case complements the above two extreme cases, which is when the risk assumes a finite value with some intermediate confidence level.

%%%%%%%%%%%%%%%%%%%%%%%%%%%%%%%%%%%%%%%%%%%%%%%%%%%%%%%%%%%%%%%%%%%%%%%%%%%%%%%%%%%%%
\section{Risk of Cascading Collisions in Special Graph Topologies}\label{specialgraph}

The topology of the underlying communication graph plays an essential role in noise  propagation in a platoon  \cite{siami2016fundamental}, which it will naturally affect the risk of cascading collisions. In this and the next sections, we investigate several graph typologies with certain symmetries to understand the behavior of the cascading risk w.r.t the spatial locations of prior collisions.

In some applications, such as platoon of robots in a lab environment that is equipped by a Motion Capture System for localization purposes, all messages (as wireless signals) are broadcast to all robots. In such cases, the underlying communication graphs are all-to-all connected and complete. To this end, let us consider an unweighted complete communication graph, i.e., $k_{i,j} = 1$ for all $i,j \in \mathcal{V}$. The eigenvalues of the corresponding Laplacian matrix are: $\lambda_1 = 0$ and $\lambda_2 = \ldots = \lambda_n = n$. 

\begin{lemma} \label{lem:sig_complete}
    For a platoon of vehicles with a  complete communication graph, the vector of steady-state inter-vehicle distances is $\bm{\bar{d}} \sim \mathcal{N}(d \bm{1}_n, \Sigma)$ and its covariance matrix $\Sigma$ is defined element-wise by
    \[
        \sigma_{ij}: = 
            \begin{cases}
                \sigma_c,               &\text{ if} ~ i = j\\
                -\frac{1}{2}\sigma_c,    &\text{ if} ~ |i-j| = 1\\
                0,                      &\text{ if} ~ |i-j| > 1
            \end{cases},
    \]
    where $\sigma_c = \frac{g^2 \tau^3 f(n \tau, \beta \tau)}{\pi}$ for all $i, j = 1,\dots,n-1$.
\end{lemma}

\begin{proof}
    Following Lemma \ref{lem:d_steady}, since the eigenvalues $\lambda_k$ are identical for all $k \in \{2,\dots,n\}$, one has
    $$
        \sigma_{ij} = g^2 \frac{\tau^3}{2\pi} \; f(n \tau, \beta \tau) \sum_{k=2}^{n} \phi(i,j,k),
    $$
    in which $q_{i,j}$ denotes the element of the matrix $Q = [\bm{q}_1 | \dots | \bm{q}_n]$, and $$\phi(i,j,k) = ( q_{k,i+1}q_{k,j+1} - q_{k,i+1}q_{k,j} - q_{k,i}q_{k,j+1} + q_{k,i}q_{k,j} ).$$ From the orthogonality property of the normalized eigenvectors, it is known that $\sum_{k} q_{k,i} q_{k,j} = \delta_{ij}$, in which $\delta_{ij}$ denotes the Kronecker delta. We have if $i = j$, then
    \begin{align*}
        \sigma_{ij} = g^2 \frac{\tau^3}{2\pi}  f(n \tau, \beta \tau) \big( 1 - 0 - 0 + 1\big) = \frac{1}{\pi} f(n \tau, \beta \tau) g^2  \tau^3.
    \end{align*}
    Using the same line of arguments, one has for $|i-j| = 1$, which is equivalent to $j = i+1$ or $i = j+1$,
    $$
        \sigma_{ij} 
        % =  g^2 \frac{\tau^3}{2\pi} \; f \bigg( 0 - 1 - 0 + 0 \bigg) 
        = -\frac{1}{2\pi} \; f(n \tau, \beta \tau) \; g^2 \; \tau^3.
    $$
    For $|i-j| > 1$, it follows immediately that 
    $
        \sigma_{ij} 
        % =  g^2 \frac{\tau^3}{2\pi} \; f \bigg( 0 - 0 - 0 + 0 \bigg) 
        = 0
    $.
\end{proof}

\begin{figure}[t]
    \centering
	\includegraphics[width=\linewidth]{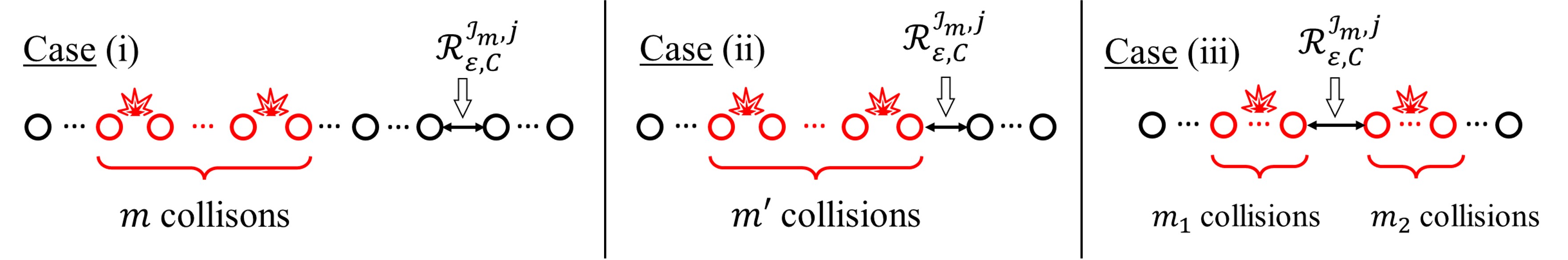}
	\caption{This figure depicts all three cases for the relative location between prior collisions and the pair of interest. }
	\label{fig:cases}
\end{figure}

The result of this lemma asserts that the corresponding covariance matrix of the all-to-all connected platoon has a symmetric tri-diagonal structure. The spatial location of the prior collisions $\mathcal{I}_m$ with respect to the $j$'th pair  creates three possible scenarios: (i) the $j$'th pair is not adjacent to any of the prior collisions, (ii) the $j$'th pair is only adjacent to $m'$ consecutive collisions on one side, and (iii) the $j$'th pair is surrounded by $m_1$ and $m_2$ consecutive collisions. These three cases are illustrated in Fig. \ref{fig:cases}. 

For the exposition of the following result, let us consider a $(\hat{m}+1) \times (\hat{m}+1)$ covariance matrix $\hat{\Sigma}$ as a simplified version of \eqref{eq:new-sig}, but with only the adjacent collisions of size $\hat{m}$. For case (ii), one has $\hat{m} = m'$ and for case (iii) we have $\hat{m} = m_1+m_2$. The inverse of $\hat{\Sigma}_{22}$ is denoted by \linebreak $\hat{\Sigma}_{22}^{-1} := [\alpha_{i,j}]$. 
\begin{figure*}[t!]
    \begin{subfigure}[t]{.24\linewidth}
        \centering
    	\includegraphics[width=\linewidth]{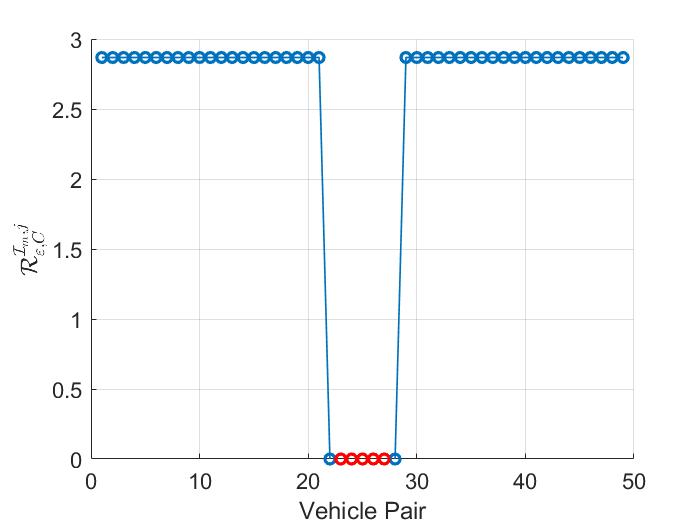}
    	\caption{The complete graph.}
    \end{subfigure}
    \hfill
    \begin{subfigure}[t]{.24\linewidth}
        \centering
    	\includegraphics[width=\linewidth]{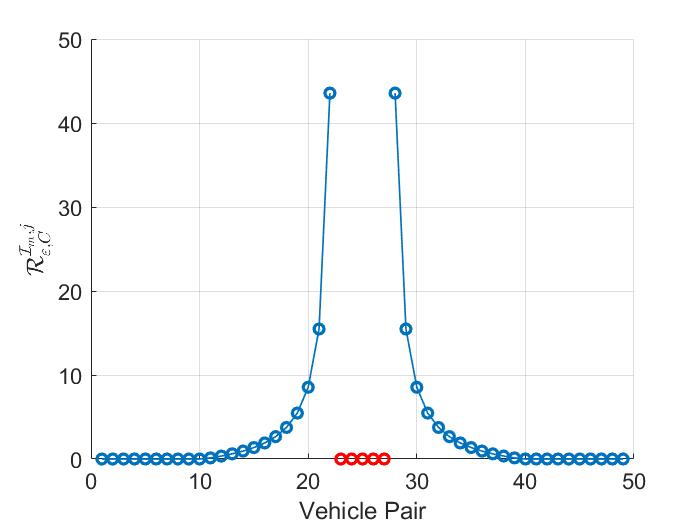}
    	\caption{The path graph.}
    \end{subfigure}
    \hfill
    \begin{subfigure}[t]{.24\linewidth}
        \centering
    	\includegraphics[width=\linewidth]{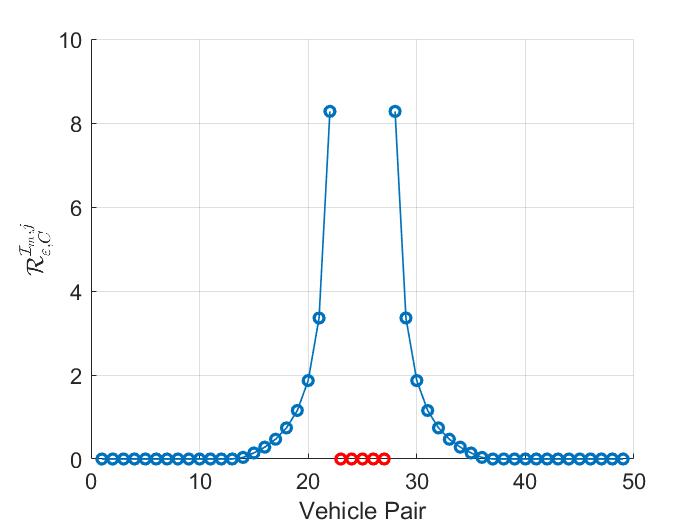}
    	\caption{The 1-cycle graph.}
    \end{subfigure}
    \hfill
    \begin{subfigure}[t]{.24\linewidth}
        \centering
    	\includegraphics[width=\linewidth]{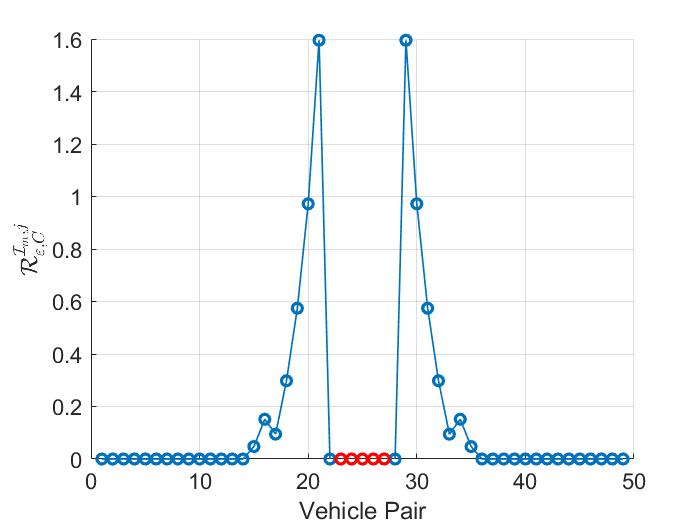}
    	\caption{The 5-cycle graph.}
    \end{subfigure}
    \caption{The risk profile $\bm{\mathcal{R}}^{\mathcal{I}_m}_{\varepsilon,C}$ (blue), assuming failures occur at pairs $\{23,24,25,26,27\}$ (red).}
    \label{fig:risk_collision}
\end{figure*}

\begin{theorem}     \label{thm:2}
    Suppose that the network of vehicles forms a platoon. If pairs with labels $i_1, \cdots, i_m$ have collided, then the risk of cascading collision of the $j$'th  pair is given by \eqref{eq:thm-1}, in which $\tilde{\mu}$ and $\tilde{\sigma}$ are computed as follows.
    
    \vspace{0.1cm}
    \noindent \underline{Case (i)}: In this case, we have $|i_k-j| > 1$ for all $i_k \in \mathcal{I}_m$, $\tilde{\mu} = d$, and $\tilde{\sigma} = \sqrt{\sigma_c}$. The cascading risk falls into the naive collision risk
    \begin{align} \label{eq:naive-risk}
        \mathcal{R}^{C,j}_{\varepsilon} = \dfrac{d}{\sqrt{2} \io \sigma_{j} + d} - c,
    \end{align}
    which is originally defined in \cite{Somarakis2020b}.
    
    \vspace{0.1cm}
    \noindent\underline{Case (ii)}: In this case, there exist only one $|i_k-j| = 1$ for all $i_k \in \mathcal{I}_{m'}$. Thus, one has
    $$
    \tilde{\sigma} = \sqrt{\sigj^2 - \frac{\sigma_c m'}{2(m'+1)}},
    ~
    \text{ and }
    ~ \tilde{\mu} = d + \frac{\sigma_c}{2} \sum_{k=1}^{m'} \alpha_{1,k} (d_{c_k} -  d),
    $$
    where $d_{c_k}$ is the corresponding element in $\bm{d}_c$.
    
    \vspace{0.1cm}
    \noindent \underline{Case (iii)}: In this case, there exists $k = j - 1$ and $k' = j + 1$ for which $k \in \mathcal{I}_{m_1}$ and $k' \in \mathcal{I}_{m_2}$. Thus, one gets
    $$
    \tilde{\sigma} = \sqrt{\sigj^2 - \frac{\sigma_c}{2} \frac{4m_1m_2 + m_1 + m_2}{m_1 + m_2 + 1}}
    $$
    and
    $$
    \tilde{\mu} = d + \frac{\sigma_c}{2} \sum_{k=1}^{m_1+m_2} (\alpha_{m_1,k} + \alpha_{m_1+1, k} )(d_{c_k} -  d).
    $$ 
\end{theorem}

\begin{proof}
    Since the structure of $\tilde{\Sigma}_{22}$ is a tridiagonal (Jacobi) matrix in the complete graph, we can use the result from \cite{usmani1994inversion} to compute its inverse. Let us denote $\theta_{k} = 2^{-k} \sigma_c^k (k+1)$ for $k = 1,...,m$, the inversion $\tilde{\Sigma}_{22}^{-1}: = [\alpha_{i,j}]$ is given by:
    \begin{equation}    \label{eq:inverse}
        \alpha_{ij}=\begin{cases}
        \left(\prod_{i}^{j-1} \frac{\sigma_c}{2}\right) \frac{\theta_{i-1}\theta_{m-j}}{\theta_{m}}, & i < j\\[5pt]
        \frac{2\, i \, (m+1-i)}{\sigma_c\, (m+1)}, & i = j\\[5pt]
        \left(\prod_{j+1}^{i} \frac{\sigma_c}{2}\right) \frac{\theta_{j-1}\theta_{m-i}}{\theta_{m}} , & i > j\\
        \end{cases}.
    \end{equation}
    
    \underline{Case 1}: The result is immediate since $\tilde{\Sigma}_{12} = \tilde{\Sigma}_{21}^T = [0, \dots, 0]$, the conditional mean and variance remain unchanged.
    
    \underline{Case 2}: Given the result from Case (i), since the existing failures have no impact for the non-adjacent vehicles pair, we only consider the adjacent failure group with size $m'$. Given that $\hat{\Sigma}_{12} = \hat{\Sigma}_{21}^T = [-\sigma_c/2,0,\dots,0]$ or $[0,\dots,0,-\sigma_c/2]$. Using the symmetry, the conditional variance can be written as
    $$
        \hat{\sigma}^2 
        = \sigj^2 - \frac{\sigma^2_c}{4} \alpha_{11} = \sigj^2 - \frac{\sigma^2_c}{4} \alpha_{m'm'}
        = \sigj^2 - \frac{\sigma_c \, m'}{2(m'+1)},
    $$
    and the value of $\hat{\mu}$ can be obtained with the same lines of argument.
    
    \underline{Case 3}: In this case, let us assume labels of two groups of failures are denoted by $\mathcal{I}_{m_1} = \{i_1,..., i_{m_1}\}$ and $\mathcal{I}_{m_2} = \{i_{m_1+1},..., i_{m_1+m_2}\}$. Then, let us form a new $(m_1+m_2+1) \times (m_1+m_2+1)$ matrix $\tilde{\Sigma}$ as follows
    \begin{equation*}
        \tilde{\Sigma} = \begin{pmatrix}\,
        \tilde{\Sigma}_{11} &\tilde{\Sigma}_{12}\\
        \tilde{\Sigma}_{21} &\tilde{\Sigma}_{22}\,
        \end{pmatrix},
    \end{equation*}
    where
    $\tilde{\Sigma}_{11} = \sigma_{j}^2, \, \tilde{\Sigma}_{12} = \tilde{\Sigma}_{21}^T = [\sigma_{j i_1},...,\sigma_{j i_{m_1+m_2}}]$, and  $\tilde{\Sigma}_{22} = [\sigma_{k_1 k_2}]_{k_1, k_2 \in (\mathcal{I}_{m_1}\bigcup\mathcal{I}_{m_2})} \in \R^{(m_1+m_2) \times (m_1+m_2)}$. Then, by using the result obtained from \eqref{eq:inverse}, one can obtain the result using the same lines of argument.
\end{proof}

\begin{figure}
    \begin{subfigure}[t]{.49\linewidth}
        \centering
    	\includegraphics[width=\linewidth]{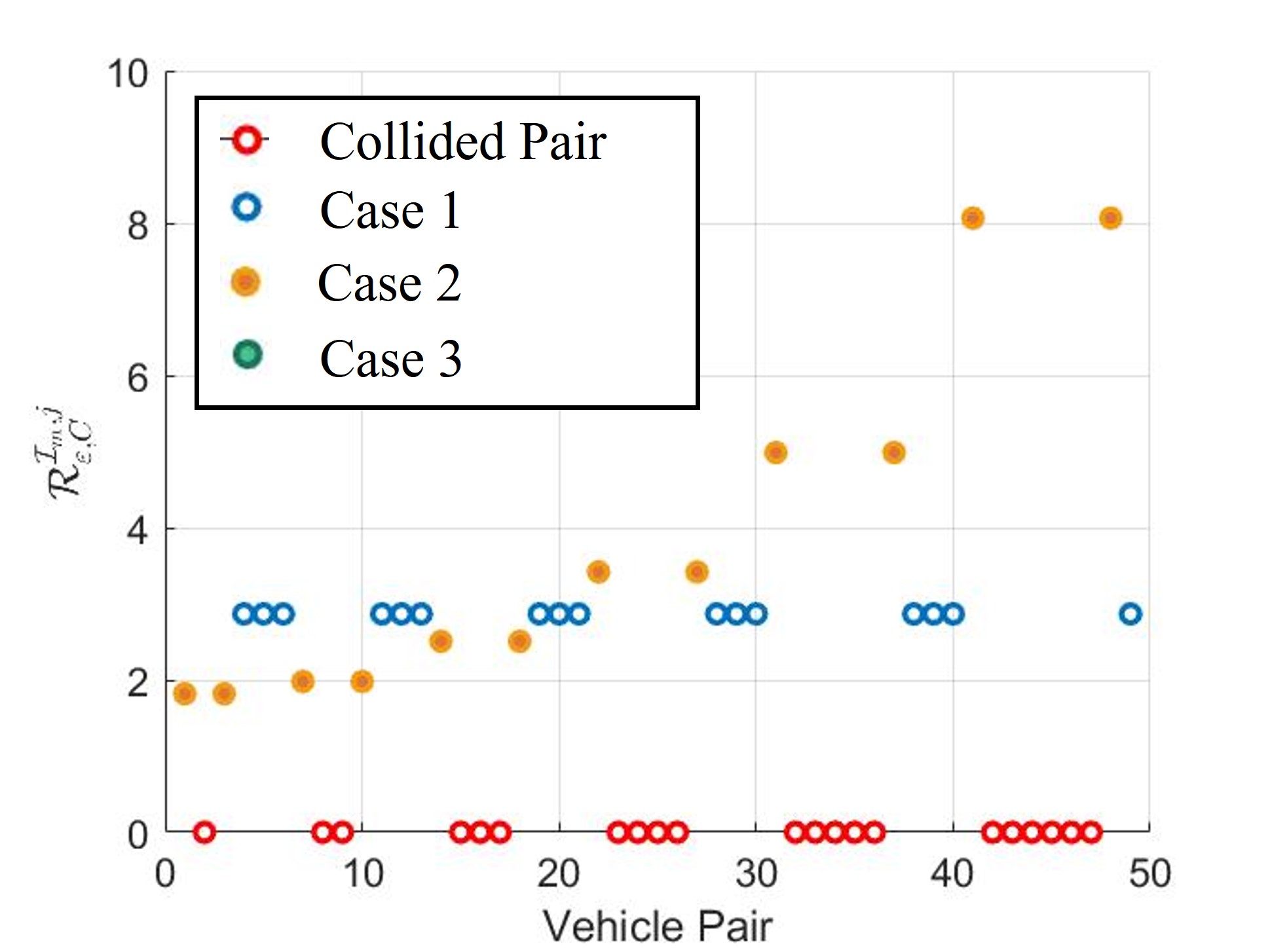}
    	\caption{The risk profile when there exists case (i) and case (ii) in Theorem \ref{thm:2}.}
    \end{subfigure}
    \hfill
    \begin{subfigure}[t]{.49\linewidth}
        \centering
    	\includegraphics[width=\linewidth]{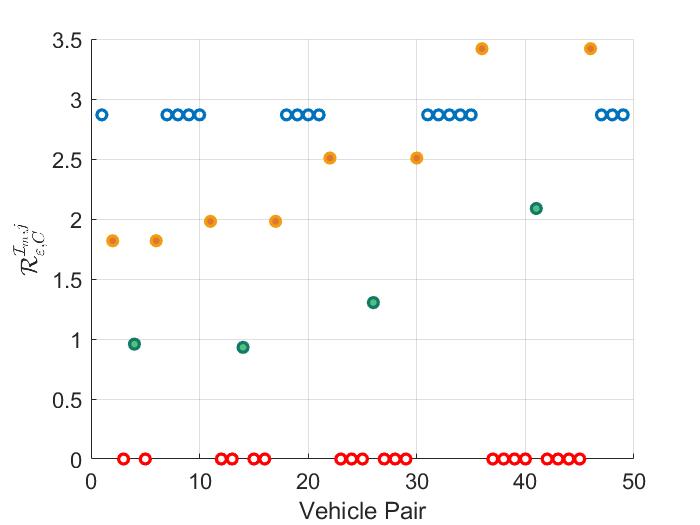}
    	\caption{The risk profile when there exists case (i), case (ii), and case (iii) in Theorem \ref{thm:2}.}
    \end{subfigure}
    \caption{The risk profile $\bm{\mathcal{R}}^{\mathcal{I}_m}_{\varepsilon,C}$ with a complete graph. For a better presentation of the result, we assume all the failed vehicle pairs are found at $\bm{d}_{c_k} = 1.1d$. The cascading risk $\casr$ obtains the same value with $\mathcal{R}^{C,j}_{\varepsilon}$ \cite{Somarakis2020b} when it is not adjacent to any existing systemic failures. The risk values of Case (ii) and Case (iii) also behave as presented in Theorem \ref{thm:2}.}
    \label{fig:thm2}
\end{figure}

The above theorem asserts that the risk in a complete graph varies based on the relative position between the prior collisions and the pair of interest. If the pair of interest is not adjacent to a group of prior collisions, e.g., case (i), its risk level will remain the same as the naive collision risk \eqref{eq:naive-risk} as the cross-correlation vanishes when the pairs are not adjacent. This finding helps us simplify the other cases by only considering the adjacent group of collisions as the non-adjacent group has zero impact on the pair of interest. This effect is also observed in Fig. \ref{fig:thm2}.

The second and third cases follow immediately by considering if the pair of interest is adjacent to only one or two ``group(s)" of collisions. It should be emphasized that in Case (ii), the risk value $\casr$ remains the same regardless of being located at the front or the back of the group $\mathcal{I}_{m}$. By letting $m_1 = 0$ or $m_2 = 0$, the expression of case (iii) will also boil down to the one in case (ii). The simulation results for all cases are presented in Fig. \ref{fig:thm2}.

%%%%%%%%%%%%%%%%%%%%%%%%%%%%%%%%%%%%%%%%%%%%%%%%%%%%%%%%%%%%%%%%%%%%%%%%%%%%%%%%%%%%%
\section{Case Studies}  \label{sec:case-study}

% \guangyi{We have a more detailed results for all the simulations shown in the extended version, which we will cite here and publish on arXiv. } 
We consider a platooning problem with $n = 50$ vehicles. Systemic set parameters are set to $ c = 2, d = 3~ \text{and}~ \varepsilon = 0.1$, and the existing failures are considered as collisions, i.e., $\bm{d}_c = \bm{0}$, for all case studies. To measure the risk of cascading collisions among the entire platoon, let us introduce the {\it risk profile} as follows 
% \chris{Please change "failures" to "collisions" }

$$
    \bm{\mathcal{R}}^{\mathcal{I}_m}_{\varepsilon,C} = \big[\mathcal{R}^{\mathcal{I}_m,1}_{\varepsilon,C}, \dots, \mathcal{R}^{\mathcal{I}_m,n-1}_{\varepsilon,C} \big]^T,
$$
in which $\mathcal{R}^{\mathcal{I}_m,j}_{\varepsilon,C} = 0$ if $j \in \mathcal{I}_m$. 
% \nader{will we use this notation?} \guangyi{Yes, it is used in the case study.}\nader{In that case, please move this part to where it is used.} 

\subsection{Risk of Cascading Collision}

\begin{figure*}[t!]
    \begin{subfigure}[t]{.32\linewidth}
        \centering
    	\includegraphics[width=\linewidth]{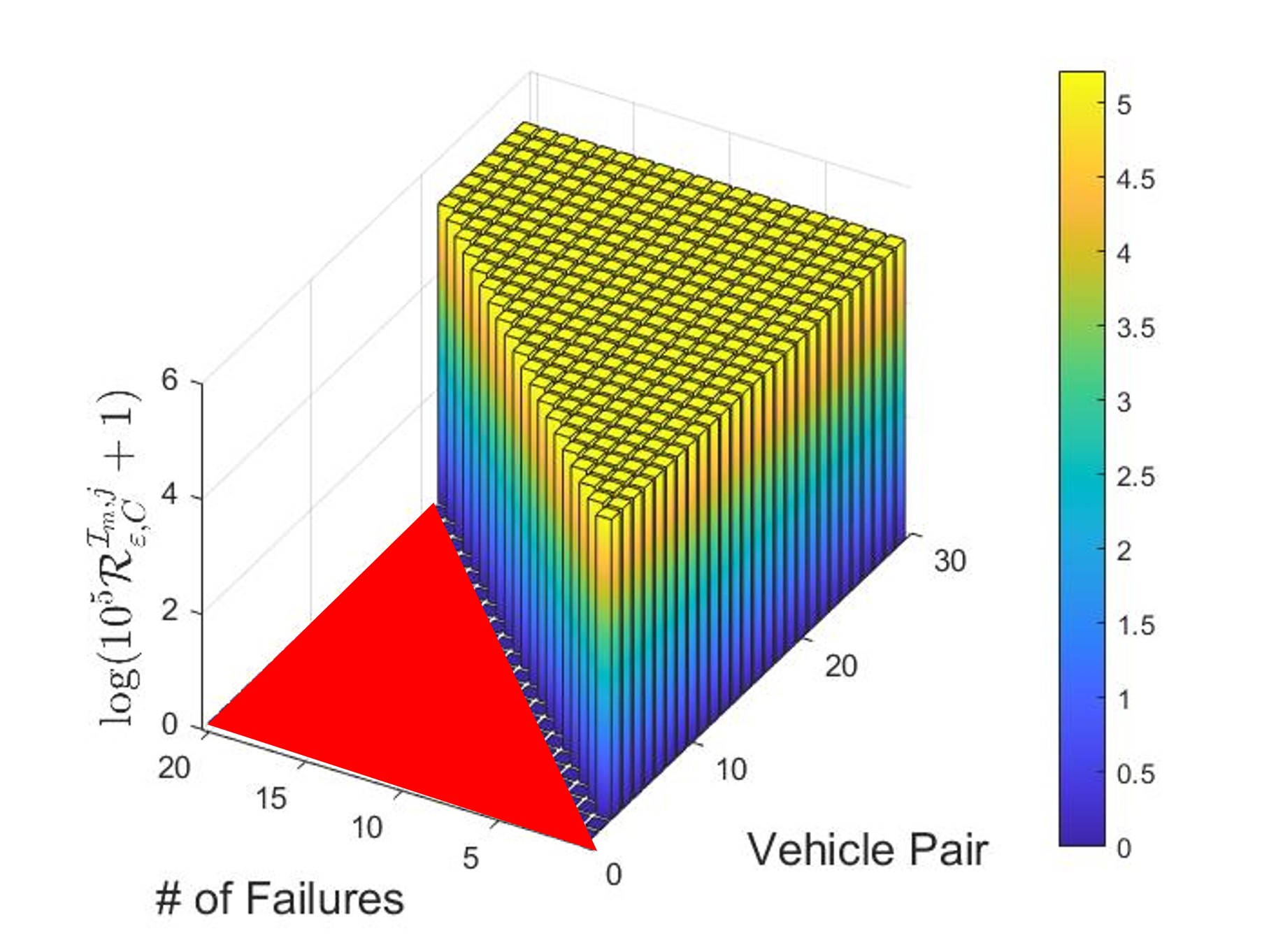}
    	\caption{The complete graph.}
    \end{subfigure}
    \hfill
    \begin{subfigure}[t]{.32\linewidth}
        \centering
    	\includegraphics[width=\linewidth]{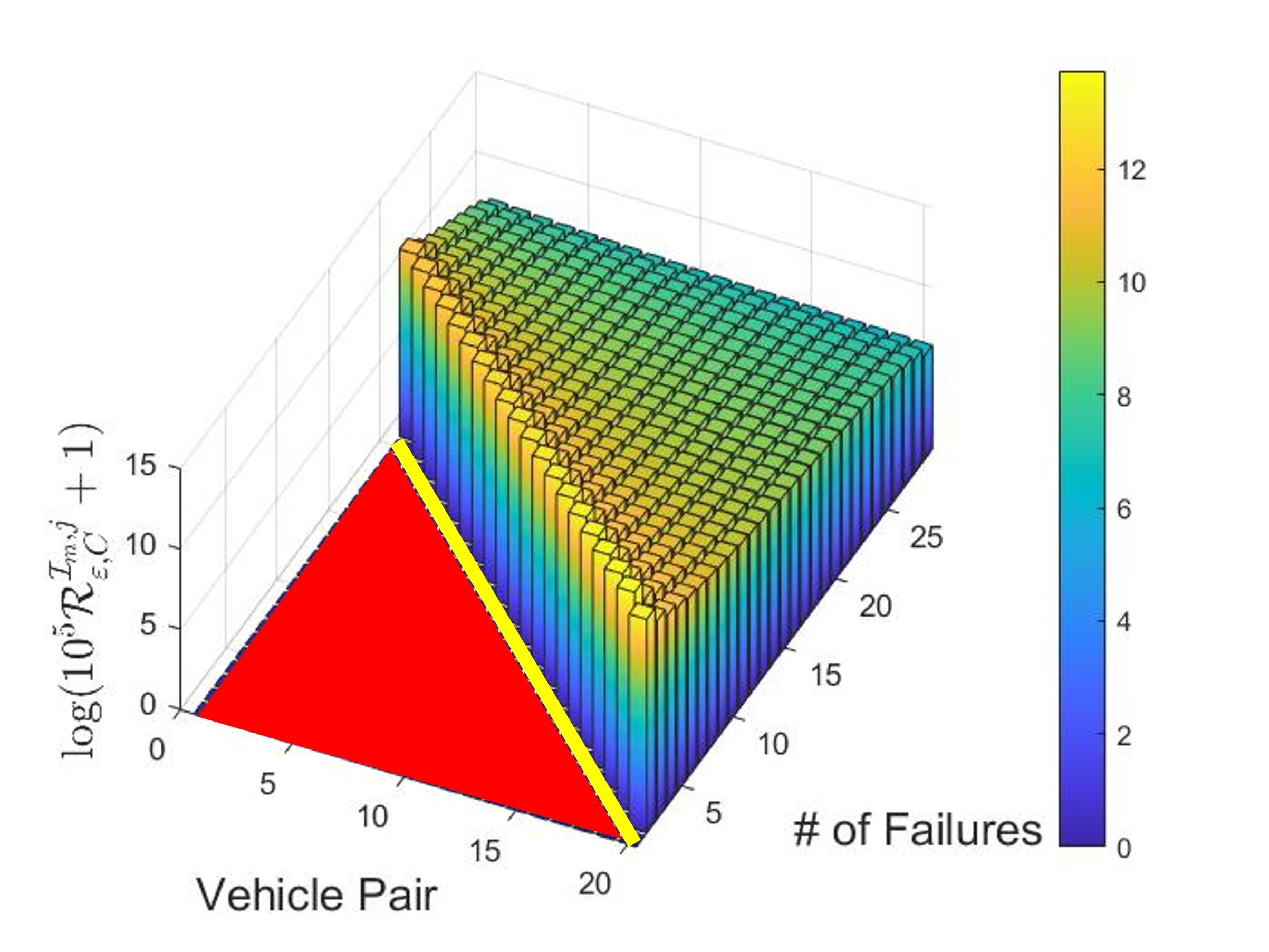}
    	\caption{The path graph.}
    \end{subfigure}
    \hfill
    \begin{subfigure}[t]{.32\linewidth}
        \centering
    	\includegraphics[width=\linewidth]{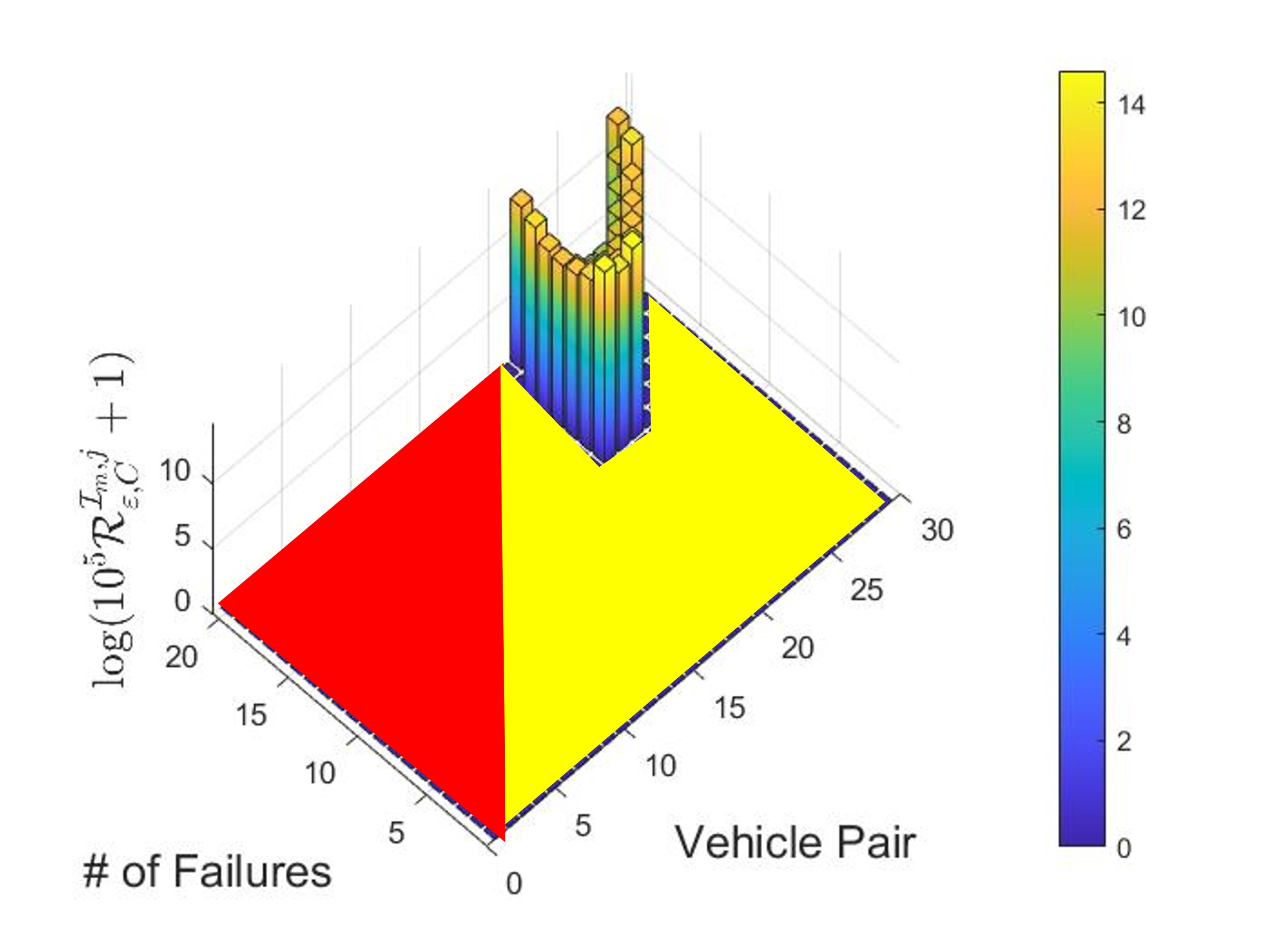}
    	\caption{The 1-cycle Graph.}
    \end{subfigure}
    \medskip
    \begin{subfigure}[t]{.32\linewidth}
        \centering
    	\includegraphics[width=\linewidth]{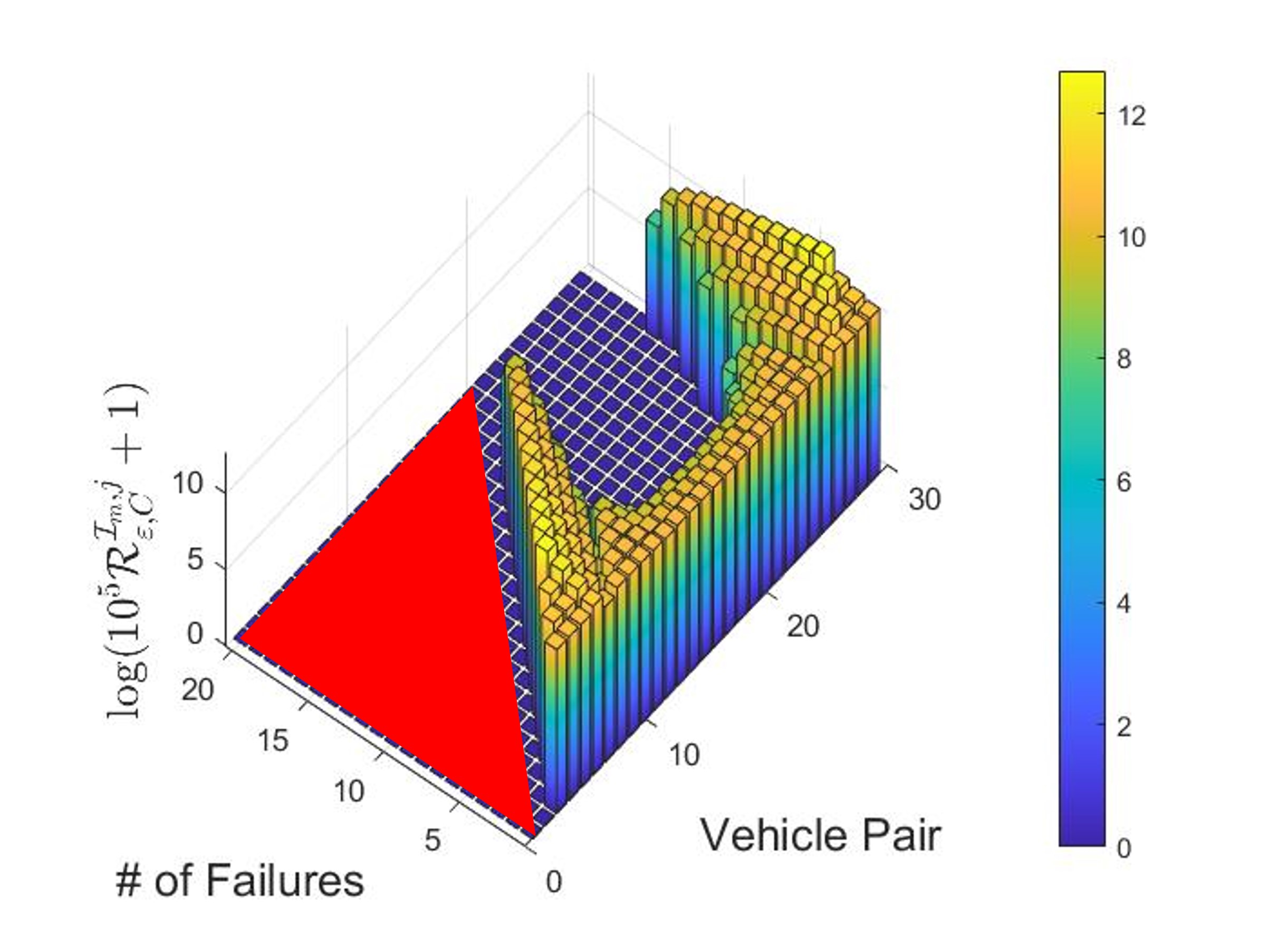}
    	\caption{The 5-cycle graph.}
    \end{subfigure}
    \hfill
    \begin{subfigure}[t]{.32\linewidth}
        \centering
    	\includegraphics[width=\linewidth]{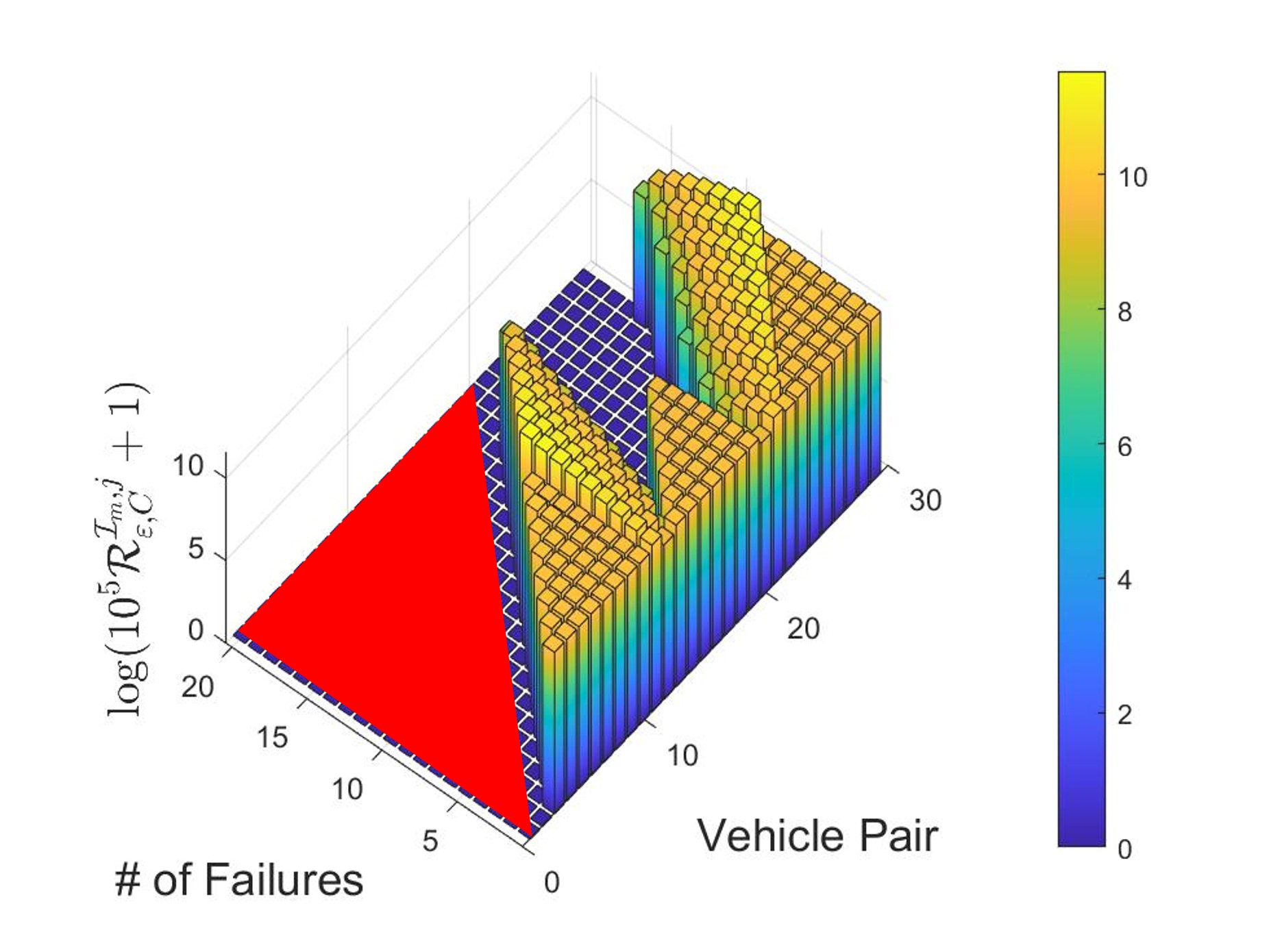}
    	\caption{The 10-cycle graph.}
    \end{subfigure}
    \hfill
    \begin{subfigure}[t]{.32\linewidth}
        \centering
    	\includegraphics[width=\linewidth]{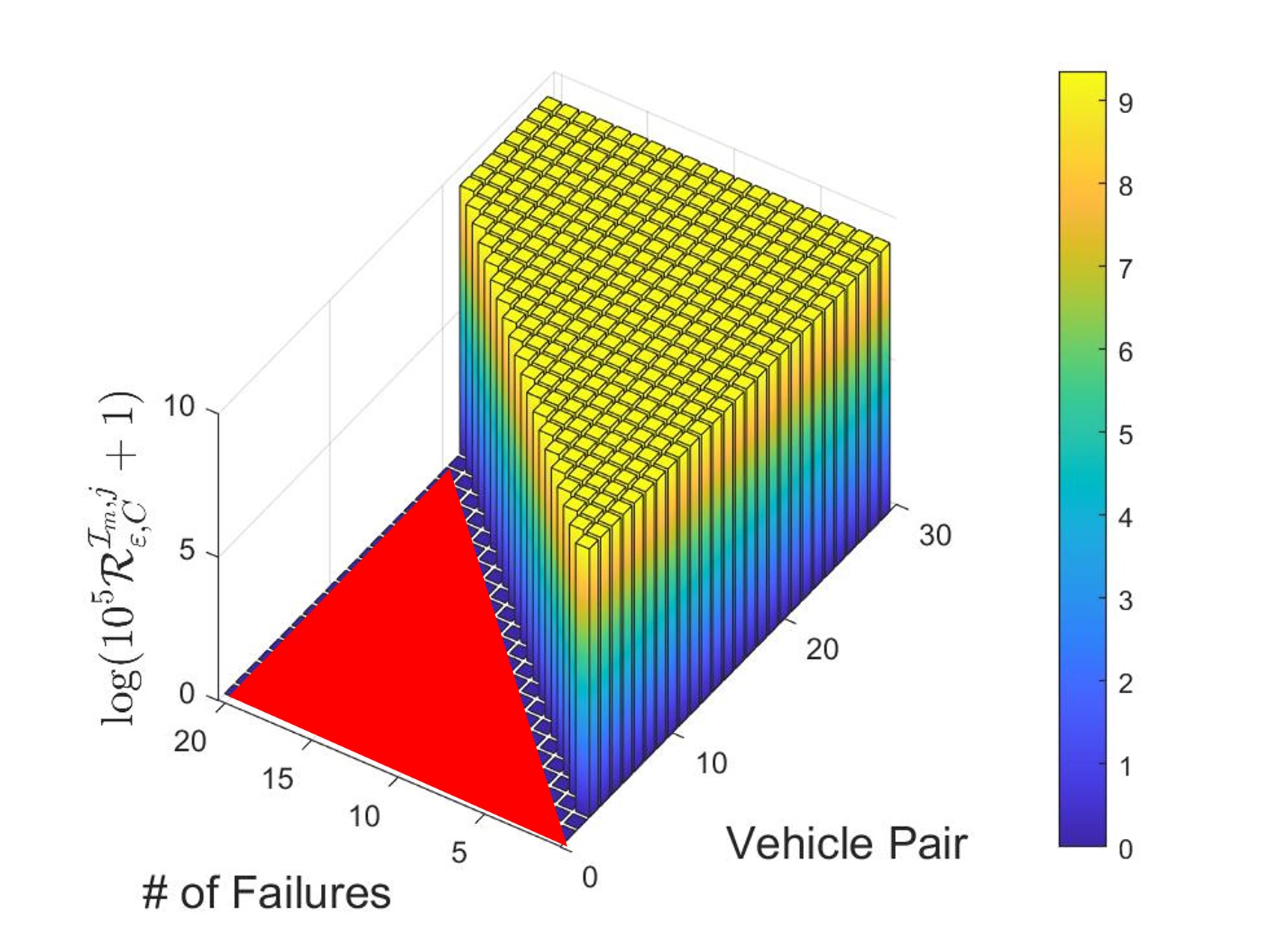}
    	\caption{The 15-cycle graph.}
    \end{subfigure}
    \caption{This figure depicts the risk profile among various failure scale and communication graphs. The red region denotes the failed vehicles and the yellow region denotes when $\casr = \infty$.}
    
    \label{fig:num_failures}
\end{figure*}

The risk profile of cascading collision $\bm{\mathcal{R}}^{\mathcal{I}_m}_{\varepsilon,C}$ for all pairs of vehicles in the platoon is evaluated with the closed-form representation derived in Theorem \ref{thm:1}. In Fig. \ref{fig:risk_collision}, we evaluate the risk profile among various communication graphs that we discuss below.

\underline{Complete Graph}: We assume the all to all communication is available, and set $g = 10, \tau = 0.03, \text{ and } \beta = 0.005$. It is shown that the vehicle pair that is adjacent to the failure group obtains a zero risk value, and the remaining pairs obtain the same value as the naive risk $\mathcal{R}^{C,j}_{\varepsilon}$ in \cite{Somarakis2020b}, which is also observed through the closed form of Eq. \eqref{eq:naive-risk}. 

\underline{Path Graph} \cite{van2010graph}: 
We assume vehicles can only communicate with their front and rear neighbors in the platoon, which can be interpreted as a car platoon on the highway. We set $g = 0.1, \tau = 0.03, \text{ and } \beta = 2$. The results indicate that the impact from the failures in a path graph will first exasperate and then dilute as the distance to the failure increases. 

\underline{p-Cycle Graph} \cite{van2010graph}: 
We assume vehicles communicate to their $p$ immediate graph neighbors, and we set $g = 0.1, \tau = 0.01, \text{ and } \beta = 2$. The behavior of the risk profile resembles the one in the path graph when using a $1-$cycle graph for communication. However, when $p=5$, the risk of the adjacent pairs are reduced to $0$ which indicates it is relatively safe to stay close to the existing failures with the current setting. 

% As the value of $p$ increases, the risk profile pattern resembles the one in a complete graph.

\subsection{Impacts from the Characteristics of Collisions}
The existing collisions affect the cascading risk in a vehicle platoon in a compound manner since the failure's dimension has been lifted from one to $m$. Hence, one should expect the risk of cascading collisions to be affected by the scale of existing failures, the relative position of failures, and the communication graph topology. 

\subsubsection{Scale of Failures}
It is instinctive to notice that the number of collisions affects $\casr$ since the statistics of the conditional distribution $\bar{d}_j$ is highly dependent on the value of $m$. To reveal this relation, we assume there exists $1, \dots, 20$ failures at the beginning of the platoon and evaluate the risk profile $\bm{\mathcal{R}}^{\mathcal{I}_m}_{\varepsilon,C}$ over different communication graph topologies, which is depicted in Fig. \ref{fig:num_failures}. There are a few key observations that are worth reporting. First, in a complete graph, the pair that is adjacent to the failure group $\mathcal{I}_m$ obtains the lowest risk value, and the risk of cascading collision for the other vehicles remains identical to the naive risk \eqref{eq:naive-risk} when the scale of existing collisions changes. This result implies the complete graph is equally endurant to both small and large scale collisions. Second, in the path graph, the most dangerous pair always occur subsequent to the existing failures (e.g., $\casr = \infty$ for all $j$ that is adjacent to $\mathcal{I}_m$, as shown in in Fig. \ref{fig:num_failures} (b).),  suggesting the collision is very likely to cascade in this setting, and a path graph is not an optimal selection for communication in the view of collisions. Finally, in a $p-$cycle graph, the value of $\casr$ decreases as the scale of failures increases, indicating that the platoon using $p-$cycle graph is more durable to the large scales collisions.

\subsubsection{Sparsity of Failures}
\begin{figure}[t]
    \begin{subfigure}[t]{.49\linewidth}
        \centering
    	\includegraphics[width=\linewidth]{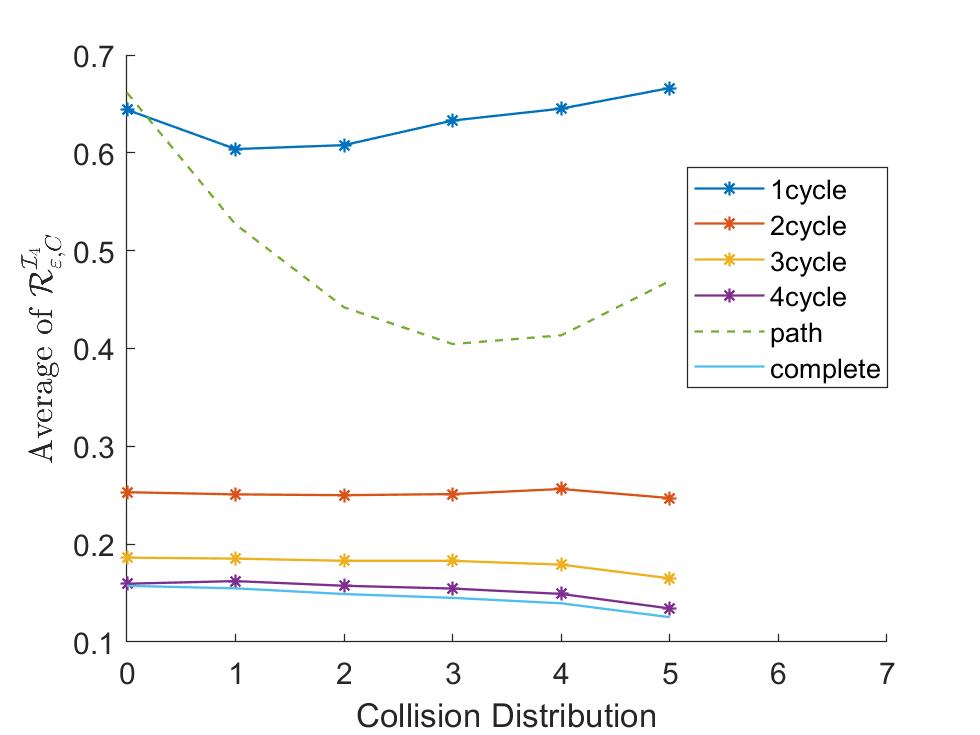}
    	\caption{$m = 5$ and $n = 10$.}
    \end{subfigure}
    \hfill
    \begin{subfigure}[t]{.49\linewidth}
        \centering
    	\includegraphics[width=\linewidth]{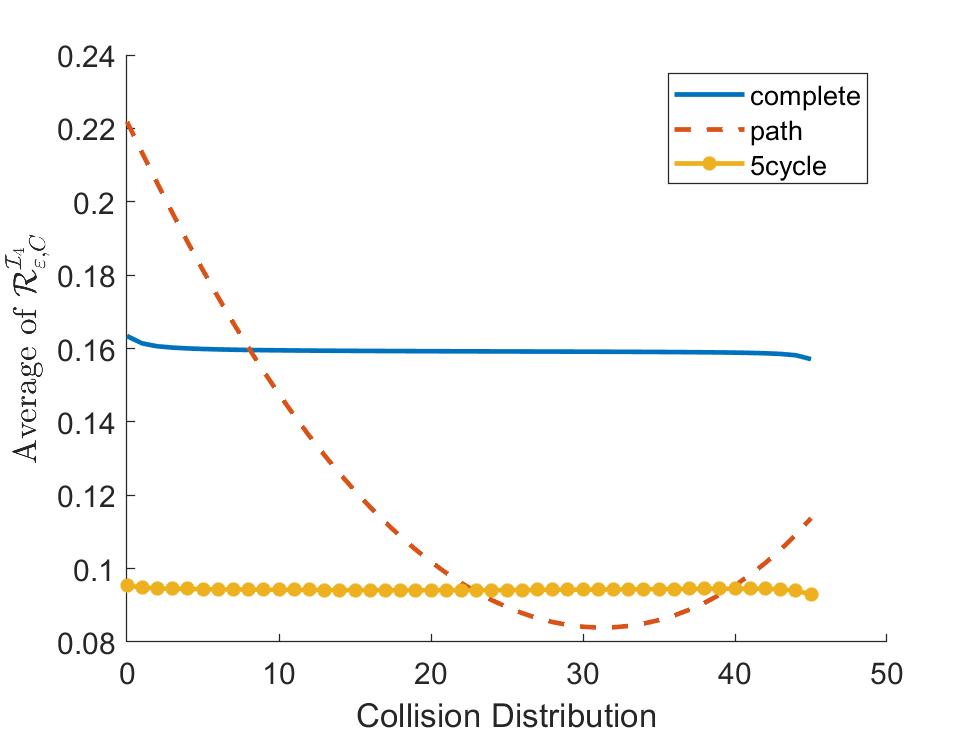}
    	\caption{$m = 5$ and $n = 50$. }
    \end{subfigure}
    \caption{Average risk of cascading collisions with various sparsity of the existing failures.}
    \label{fig:sparsity}
\end{figure}

When the scale of failures is fixed, the position distribution of existing failures also affects the risk of new collisions. For exposition of the next result, let us consider an indicator vector $\chi = [\chi_k]$, where $k = i_1, i_1+1, ..., i_m$ and $i_1,..., i_m \in \mathcal{I}_m$ are the labels of the existing failures. The value of $\chi_k$ is given by the following indicator function:
\begin{align*}
    \chi_k = \bm{1}_{\mathcal{I}_m } (k) = \begin{cases}
        1 \text{~ if } k \in \mathcal{I}_m \\
        0 \text{~ if } k \notin \mathcal{I}_m
    \end{cases}.
\end{align*}
Then, the number of zeros, i.e., the sparsity of $\chi$, represents how sparse are the positions of failures. To reveal how it will affect the cascading risk in a vehicle platoon, we assume $m = 5$ in both a $n =10$ platoon and a $n = 50$ platoon and evaluate the average risk value among all possible combinations of $\chi$\footnote{For the sparsity level $0 \leq k \leq n-m$, the total possible number of combinations is $\binom{n}{k}$.} with the same sparsity level, as depicted in Fig. \ref{fig:sparsity}.

We highlight that in most of the tested graph structures, the average risk decreases with the increase of the sparsity of $\chi$ in most cases, indicating the platoon is more vulnerable to clustered failures. Hence, in applications like highway cars or robots platoon, one wants to avoid congregate failures and keep them distributed if failures are inevitable.

\begin{figure*}[t]
    \begin{subfigure}[t]{.24\linewidth}
        \centering
    	\includegraphics[width=\linewidth]{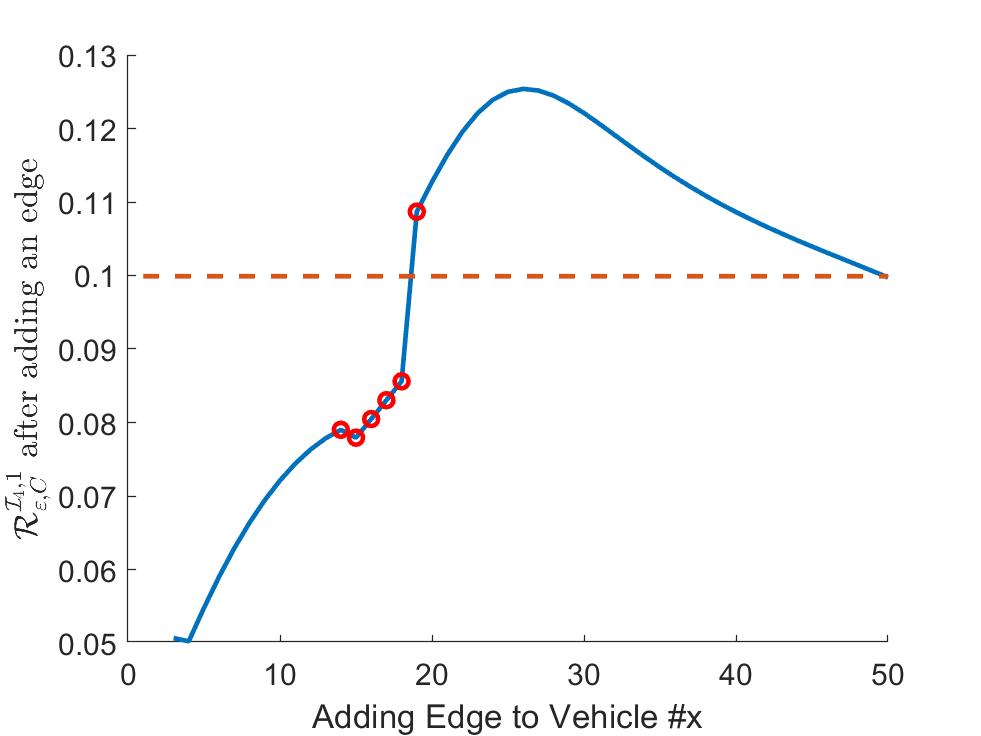}
    	\caption{The path graph with collisions occurring at pairs $\{14,15,16,17,18\}$.}
    \end{subfigure}
    \hfill
    \begin{subfigure}[t]{.24\linewidth}
        \centering
    	\includegraphics[width=\linewidth]{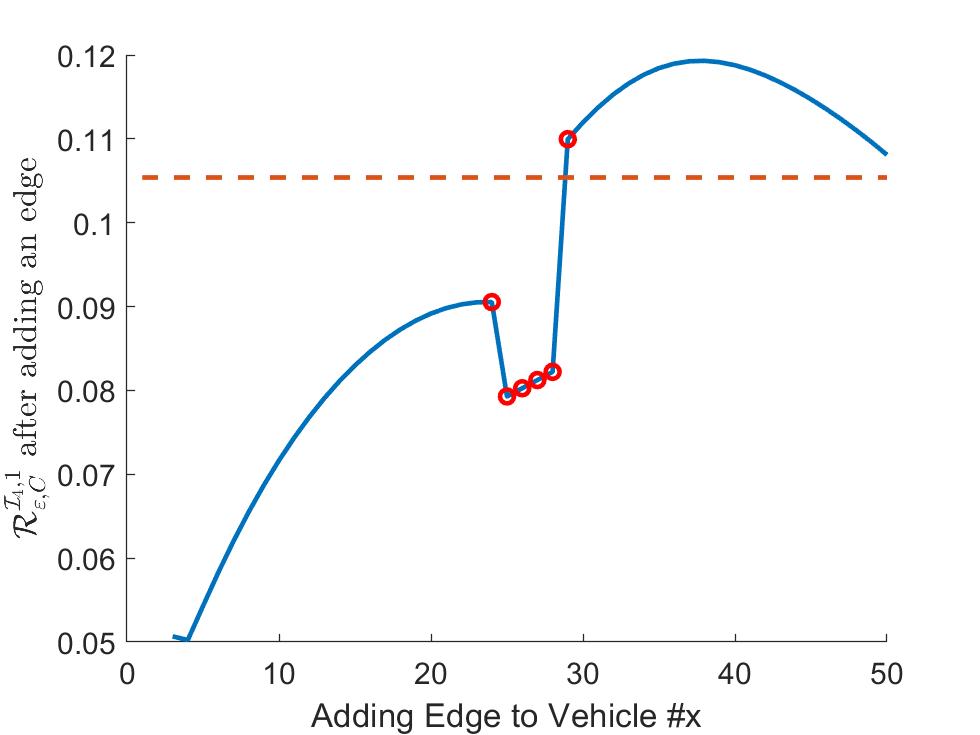}
    	\caption{The path graph with collisions occurring at pairs $\{24,25,26,27,28\}$.}
    \end{subfigure}
    \hfill
    \begin{subfigure}[t]{.24\linewidth}
        \centering
    	\includegraphics[width=\linewidth]{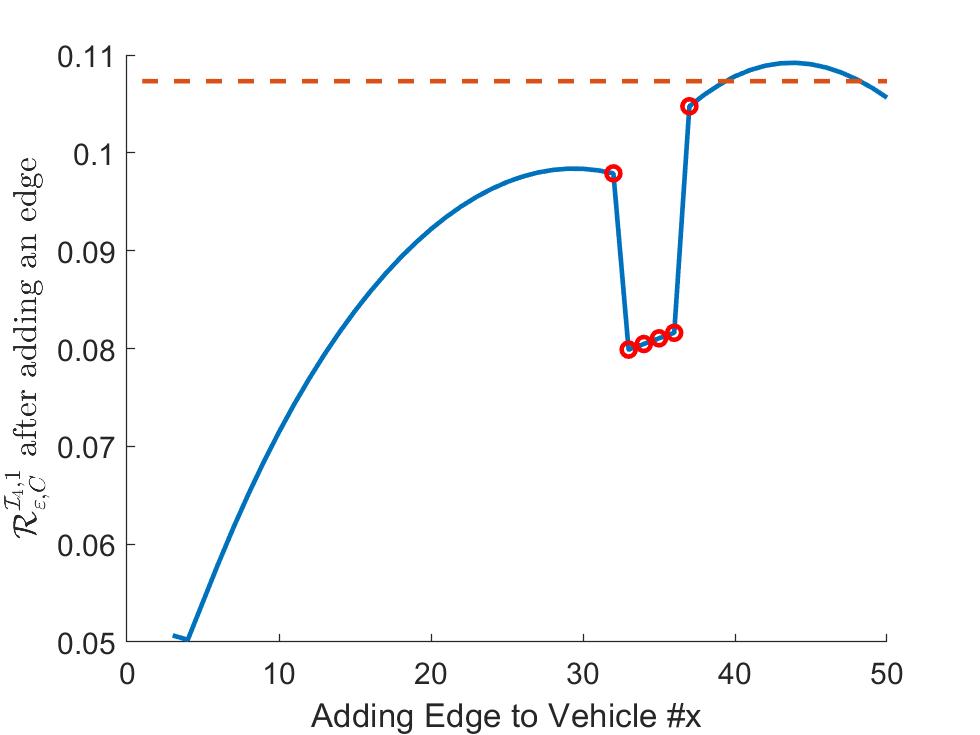}
    	\caption{The path graph with collisions occurring at pairs $\{32,33,34,35,36\}$.}
    \end{subfigure}
    \hfill
    \begin{subfigure}[t]{.24\linewidth}
        \centering
    	\includegraphics[width=\linewidth]{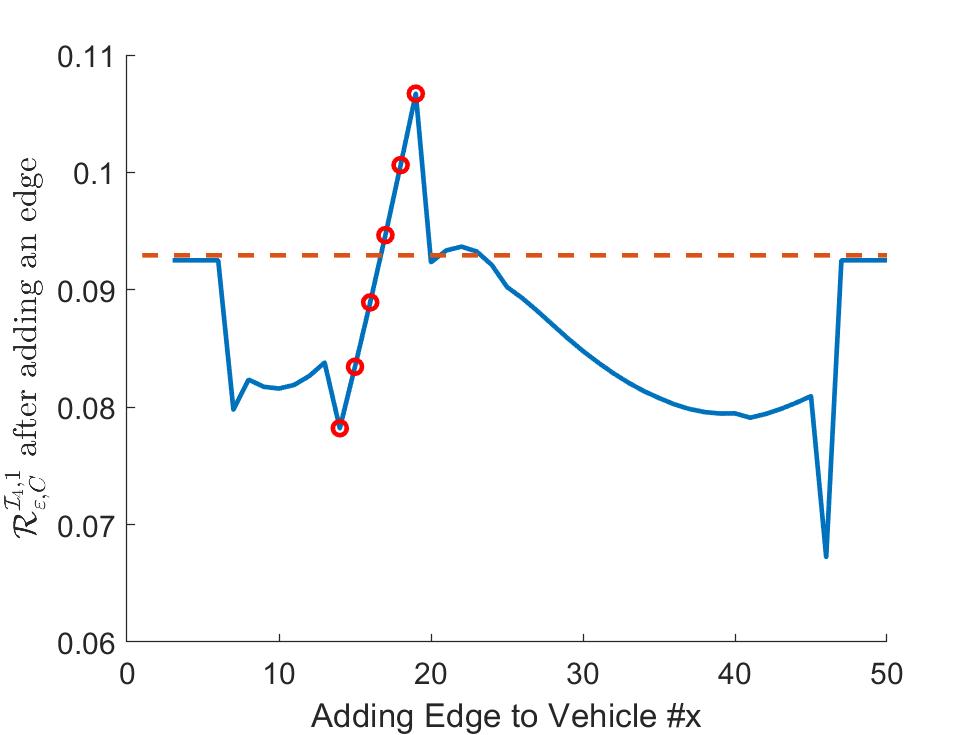}
    	\caption{$5-$cycle graph with collisions occurring at pairs $\{14,15,16,17,18\}$.}
    \end{subfigure}
    \medskip
    \begin{subfigure}[t]{.24\linewidth}
        \centering
    	\includegraphics[width=\linewidth]{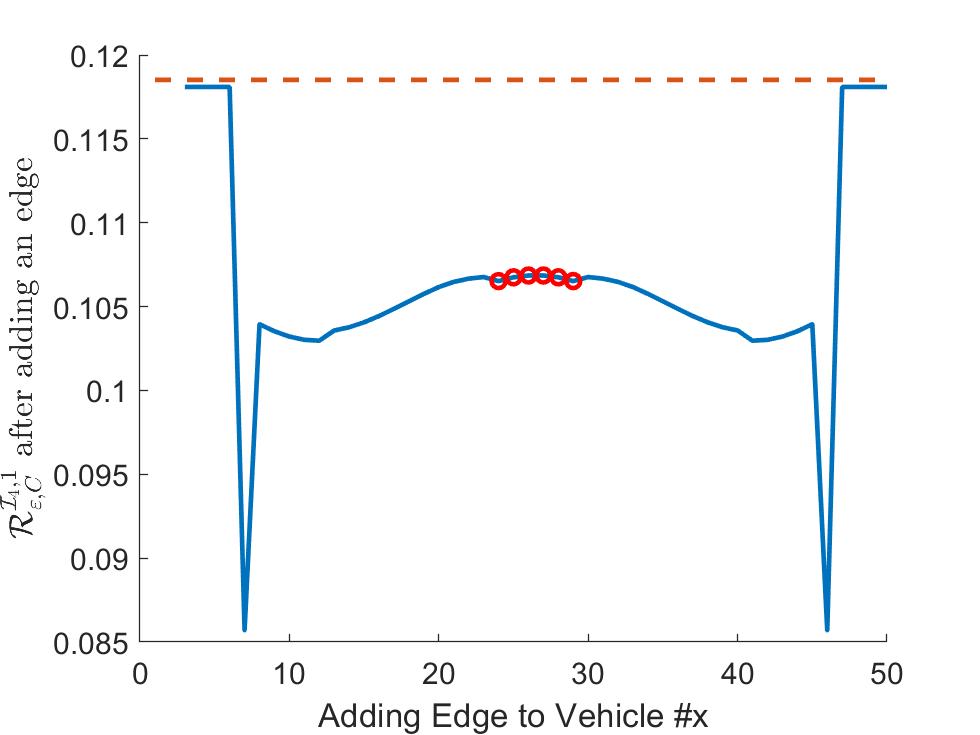}
    	\caption{$5-$cycle graph with collisions occurring at pairs $\{24,25,26,27,28\}$.}
    \end{subfigure}
    \hfill
    \begin{subfigure}[t]{.24\linewidth}
        \centering
    	\includegraphics[width=\linewidth]{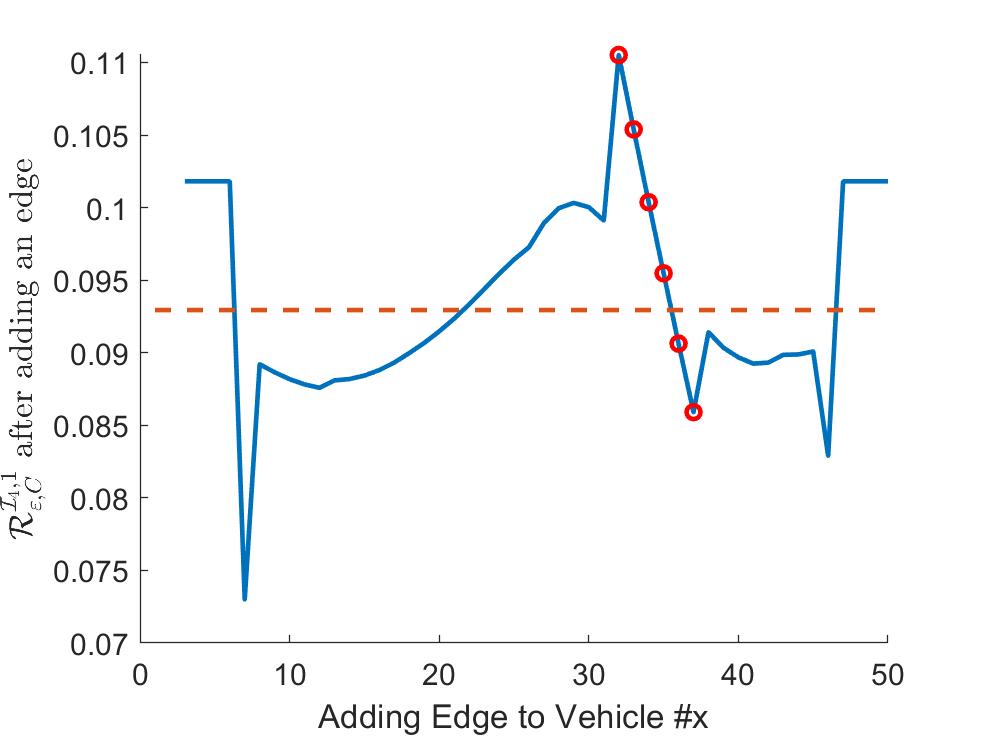}
    	\caption{$5-$cycle graph with collisions occurring at pairs $\{32,33,34,35,36\}$.}
    \end{subfigure}
    \hfill
    \begin{subfigure}[t]{.24\linewidth}
        \centering
    	\includegraphics[width=\linewidth]{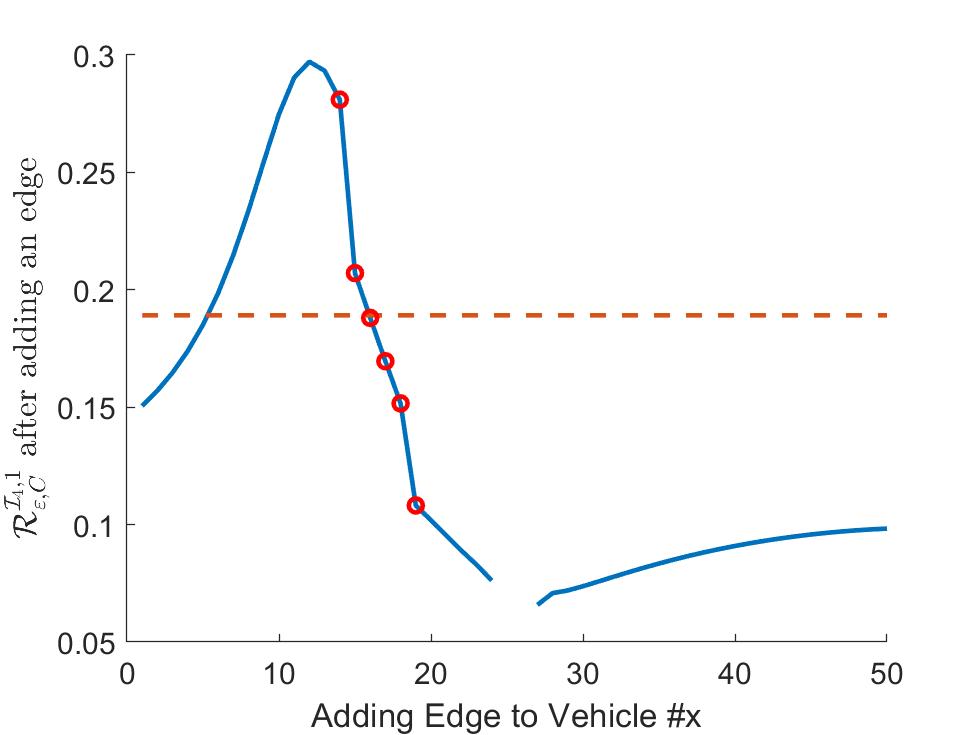}
    	\caption{The path graph with collisions occurring at pairs $\{14,15,16,17,18\}$, evaluated at the middle pair.}
    \end{subfigure}
    \hfill
    \begin{subfigure}[t]{.24\linewidth}
        \centering
    	\includegraphics[width=\linewidth]{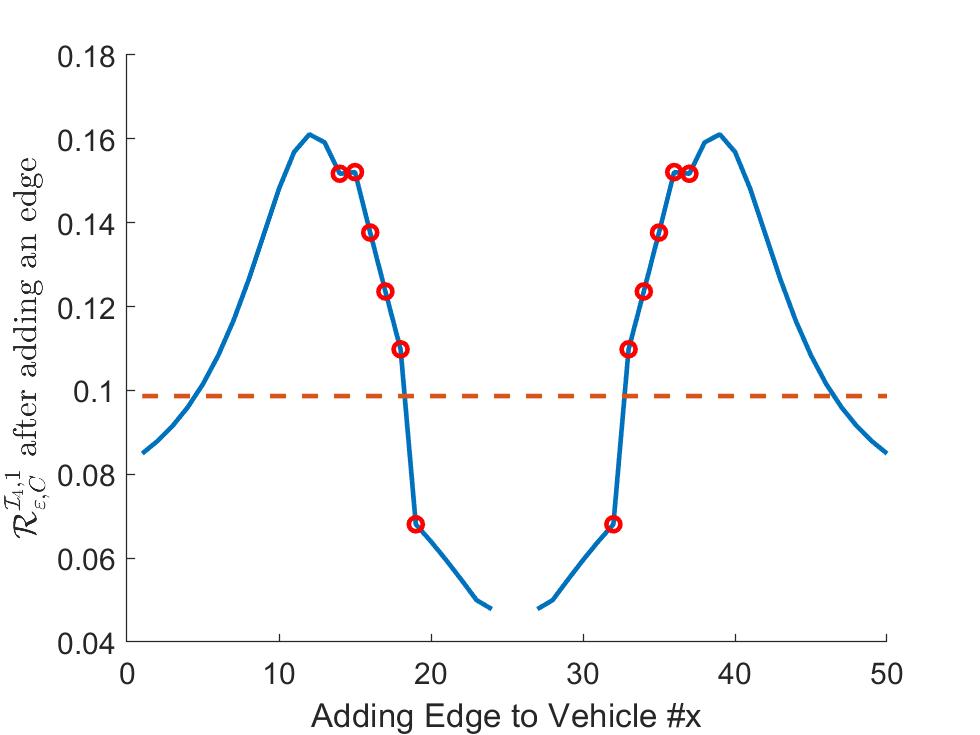}
    	\caption{The path graph with collisions occurring at pairs $\{14,15,16,17,18,32,33,34,35,36\}$, evaluated at the middle pair.}
    \end{subfigure}
    
    \medskip
    
    \begin{subfigure}[t]{.24\linewidth}
        \centering
    	\includegraphics[width=\linewidth]{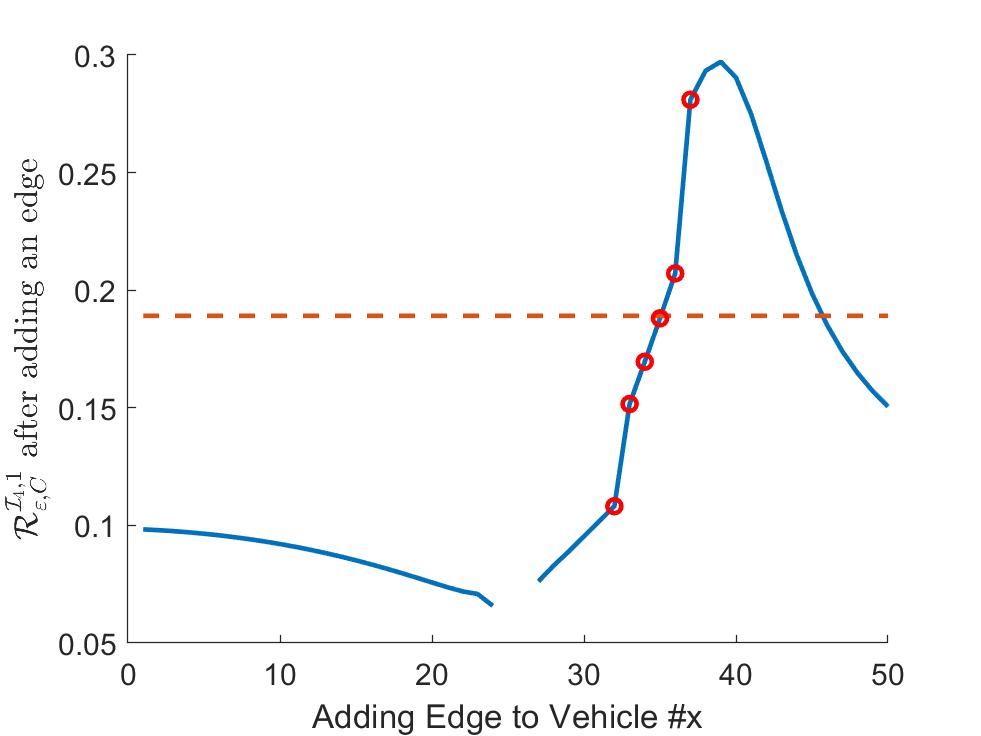}
    	\caption{The path graph with collisions occurring at pairs $\{32,33,34,35,36\}$, evaluated at the middle pair.}
    \end{subfigure}
    \hfill
    \begin{subfigure}[t]{.24\linewidth}
        \centering
    	\includegraphics[width=\linewidth]{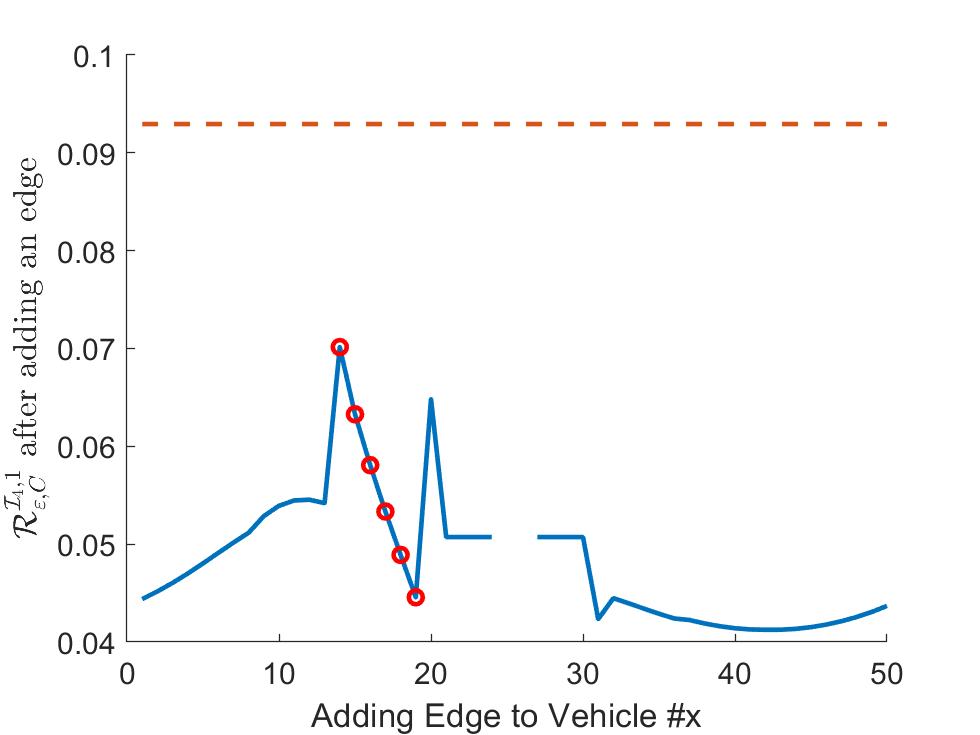}
    	\caption{$5-$cycle graph with collisions occurring at pairs $\{14,15,16,17,18\}$, evaluated at the middle pair.}
    \end{subfigure}
    \hfill
    \begin{subfigure}[t]{.24\linewidth}
        \centering
    	\includegraphics[width=\linewidth]{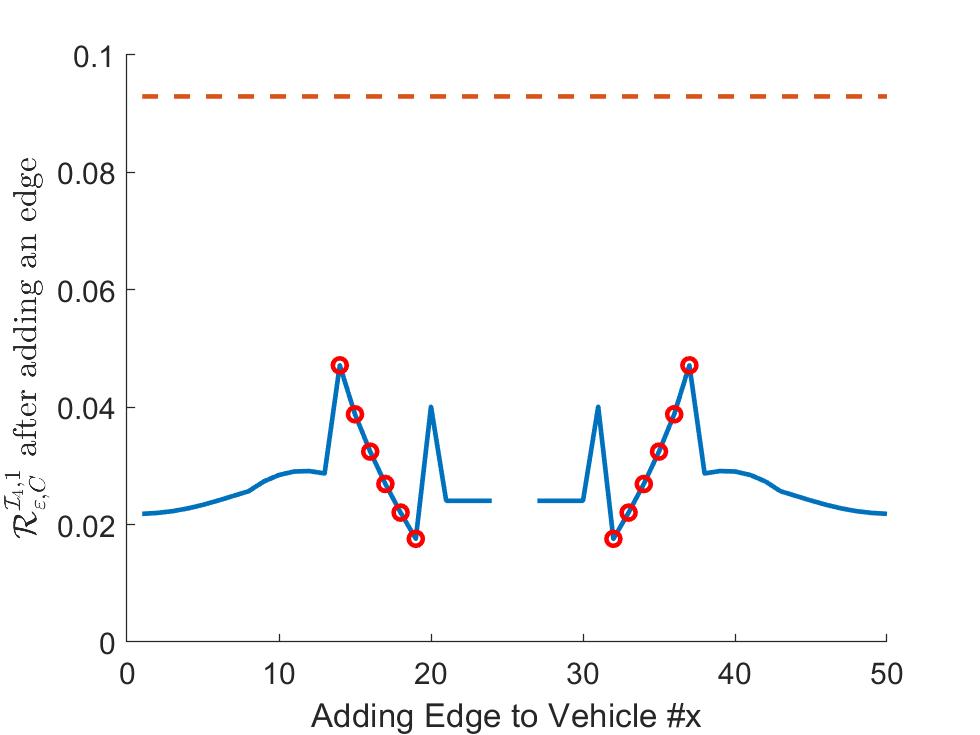}
    	\caption{$5-$cycle graph with collisions occurring at pairs $\{14,15,16,17,18,32,33,34,35,$ $36\}$, evaluated at the middle pair.}
    \end{subfigure}
    \hfill
    \begin{subfigure}[t]{.24\linewidth}
        \centering
    	\includegraphics[width=\linewidth]{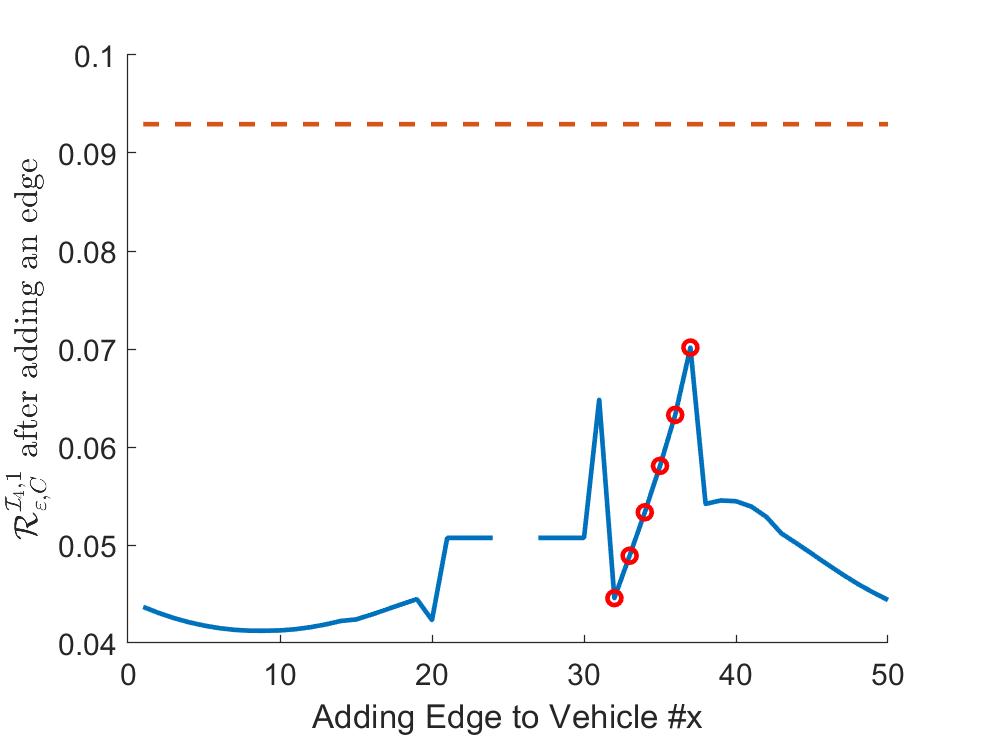}
    	\caption{$5-$cycle graph with collisions occurring at pairs $\{32,33,34,35,36\}$, evaluated at the middle pair.}
    \end{subfigure}
    \caption{ The risk profile after adding one edge to the target vehicle in the platoon. The orange dash denotes the risk level before adding the edge and the failed vehicles are marked with red dot.}
    \label{fig:add}
\end{figure*}

\subsection{Adding New Edges to the Communication Graph}
When communication capability is limited, as in most real-world problems, one may have a chance to alter the existing communication graph by adding very few new edges. It is crucial to select smartly such that they will help to reduce the risk of cascading collisions in the system. In this problem, we assume the communication graph is given, and the platoon is allowed to add two more edges\footnote{Adding two edges is more meaningful than one since the distance $\bar{d}_j$ involves two vehicles.} to the same destination in the original graph. The cascading risk is evaluated on the first and the middle pair within the platoon, as shown in Fig. \ref{fig:add}. The result suggests that, in a path graph, the risk of cascading collision is reduced if new communication is allowed with the vehicles before the collisions, and the risk deteriorates if the communication link is established to the vehicles that located on other side of the collisions in the platoon. 

% \chris{ this is not clear . Do you mean after the location of collisions ?}.  \guangyi{yes, I'm trying to find a better wording for this. It means the vehicle that locates on the other side of the collisions}

%%%%%%%%%%%%%%%%%%%%%%%%%%%%%%%%%%%%%%%%%%%%%%%%%%%%%%%%%%%%%%%%%%%%%%%%%%%%%%%%%%%%%
\section{Conclusion}
To open up the opportunity of probing into the risk of cascading failures in a multi-vehicle platoon, we develop the generalized value-at-risk measure among multiple systemic events. Using a platoon of vehicles as a motivational application, the risk profile of cascading failures (e.g., multi-vehicle accidents) is quantified using the steady-state statistics obtained from the system dynamics. With extensive simulations, we show the impact of multi-vehicle accidents with different scales and spatial distributions and comprehensively evaluate the platoon's vulnerability to cascading collisions. Some possible variations of the communication graph are also presented by adding new edges to particular graph topologies in favor of risk minimization.

% Both our theoretical findings and simulations show an evident difference to the basic version of risk of cascading failure \cite{9683468,Somarakis2020b} as the generalized risk measure proposed in this paper works with an arbitrary number of failures and it is more adaptive to the applications with the multi-vehicle system. 

%%%%%%%%%%%%%%%%%%%%%%%%%%%%%%%%%%%%%%%%%%%%%%%%%%%%%%%%%%%%%%%%%%%%%%%%%%%%%%%%%%%%%
% \appendix

% \subsection{Proofs}

% \subsubsection{Proof of Lemma \ref{lem:multi-condition}}
%  \hfill$\square$

% \subsubsection{Proof of Theorem \ref{thm:1}}

% \hfill$\square$

% \vspace{0.1cm}
% \subsubsection{Proof of Lemma \ref{lem:sig_complete}}

% \subsubsection{Proof of Theorem \ref{thm:2}}

% \begin{align*}
%     \hat{\sigma}^2 
%     &= \sigi^2 - \frac{\sigma^2_c}{4} \left(\alpha_{m_1,m_1} + 2\, \alpha_{m_1,m_1+1} + \alpha_{m_1+1,m_1+1}\right) \\
%     &= \sigi^2 - \frac{\sigma^2_c}{2} \frac{m_1^2 + 2m_1m_2 + 5m_1 + m_2}{m_1 + m_2 + 1} ,
% \end{align*}
% and one can also obtain the value for $\hat{\mu}$ with the same lines of argument.
% \hfill$\square$

%%%%%%%%%%%%%%%%%%%%%%%%%%%%%%%%%%%%%%%%%%%%%%%%%%%%%%%%%%%%%%%%%%%%%%%%%%%%%%%%%%%%%%%%%%%%%%
%END OF THE MAIN DOCUMENT
%%%%%%%%%%%%%%%%%%%%%%%%%%%%%%%%%%%%%%%%%%%%%%%%%%%%%%%%%%%%%%%%%%%%%%%%%%%%%%%%%%%%%%%%%%%%%%

\printbibliography
\end{document}